\documentclass[journal]{IEEEtran}
\usepackage{amsfonts}
\usepackage{amssymb}
\usepackage{amsmath}
\usepackage{diagbox}
\usepackage[ruled]{algorithm2e}
\usepackage{makecell} 
\usepackage{multirow}
\usepackage{xcolor}

\usepackage{graphicx}
\usepackage{cite}
\usepackage{exscale}
\usepackage{relsize}
\usepackage{algpseudocode}
\usepackage{graphics}
\usepackage{mathrsfs}
\usepackage{siunitx}
\usepackage{threeparttable}
\usepackage{url}
\usepackage{booktabs}
\usepackage{lipsum}
\usepackage{array}

\DeclareMathOperator*{\argmin}{argmin}
\newcommand{\bm}[1]{\mbox{\boldmath{$#1$}}}

\newtheorem{Rem}{Remark}

\ifCLASSOPTIONcompsoc
\usepackage[caption=false,font=normalsize,labelfont=sf,textfont=sf]{subfig}
\else
\usepackage[caption=false,font=footnotesize]{subfig}
\fi

\usepackage[T1]{fontenc}

\title{Explainable Adversarial Learning Framework on Physical Layer Key Generation Combating Malicious Reconfigurable Intelligent Surface}

\author{Zhuangkun Wei, Wenxiu Hu, Junqing Zhang, Weisi Guo, and Julie McCann

\thanks{Manuscript received xxx; revised xxx; accepted xxx. Date of publication xxx; date of current version xxx. This work is supported by the Engineering and Physical Sciences Research Council [grant number:  EP/V026763/1]. The review of this paper was coordinated by xxx. 
\textit{(Corresponding author: xxx.)}}
\thanks{Z. Wei and J. McCann are with the Department of Computing, Imperial College London, SW7 2AZ, UK. (emails: zhuangkun.wei/j.mccann@imperial.ac.uk)}
\thanks{W. Hu is with the Optoelectronics Research Centre, University of Southampton, SO17 1BJ, UK. (email: w.hu@soton.ac.uk)}
\thanks{J. Zhang is with the Department of Electrical Engineering and Electronics, the University of Liverpool, L69 3GJ, UK. (email: junqing.zhang@liverpool.ac.uk)}
\thanks{W. Guo is with the School of Aerospace, Transport, and Manufacturing, Cranfield University, MK43 0AL, UK. (email: weisi.guo@cranfield.ac.uk)}
\thanks{Color versions of one or more of the figures in this paper are available online at http://ieeexplore.ieee.org.}
	\thanks{Digital Object Identifier xxx}	
}

\begin{document}

\maketitle 

\begin{abstract}
Reconfigurable intelligent surfaces (RIS) can both help and hinder the physical layer secret key generation (PL-SKG) of communications systems. Whilst a legitimate RIS can yield beneficial impacts, including increased channel randomness to enhance PL-SKG, a malicious RIS can poison legitimate channels and crack almost all existing PL-SKGs. 
In this work, we propose an adversarial learning framework that addresses Man-in-the-middle RIS (MITM-RIS) eavesdropping which can exist between legitimate parties, namely Alice and Bob. First, the theoretical mutual information gap between legitimate pairs and MITM-RIS is deduced. From this, Alice and Bob leverage adversarial learning \textcolor{blue}{to learn a common feature space that assures no mutual information overlap with MITM-RIS.} 
Next, to explain the trained legitimate common feature generator, we aid signal processing interpretation of black-box neural networks using a symbolic explainable AI (xAI) representation. These symbolic terms of dominant neurons aid the engineering of feature designs and the validation of the learned common feature space. Simulation results show that our proposed adversarial learning- and symbolic-based PL-SKGs can achieve high key agreement rates between legitimate users, and is further resistant to an MITM-RIS Eve with the full knowledge of legitimate feature generation (NNs or formulas). This therefore paves the way to secure wireless communications with untrusted reflective devices in future 6G. 
\end{abstract}


\section{Introduction}
Wireless communication systems are susceptible to various attack vectors due to their broadcast nature. Traditional cryptography leverages the computational complexities of designed mathematical problems to generate secret keys \cite{8883127}. Such a computational burden, however, makes it less attractive, especially in securing wireless communications for massive edge users and Internet-of-Everything (IoE) devices in future 6G. 

Physical layer security (PLS) emerges as an alternative solution~\cite{6739367}. 
It relies on the unpredictable and random properties of wireless channels to secure communications. 
Physical layer secret key generation (PL-SKG) is an important branch of PLS, which exploits the random and reciprocal channel state information (CSI) between legitimate users (say Alice and Bob) to generate symmetrical secret keys \cite{7393435,zhang2020new,5422766,9963988,10368035}. The exploited reciprocal CSIs include received signal strength (RSS) \cite{7393435}, channel phases \cite{7933224}, and channel frequency responses \cite{8293762}. In these cases, Alice and Bob are required to send public pilot sequences to each other, and pursue channel estimations to acquire these reciprocal CSIs as common features, which will then be passed to the quantization  \cite{6171198,mathur2008radio}, reconciliation \cite{10.1007/3-540-48285-7_35}, and privacy amplification \cite{impagliazzo1989pseudo} modules, for key generation. 

The metric to evaluate PL-SKG is the secret key rate (SKR), which indicates the number of keys generated per common feature. Such an SKR depends on the randomness (in terms of information entropy) of common CSI features, which, however, is limited in static (quasi) channels.
To improve the SKR, extra randomness has been introduced from users' signal spaces to construct common features with high information entropy \cite{8254029,lou2017secret,7582525,7593219,8057119,8424614,7794581}. One popular method is referred to as two-way cross-multiplication \cite{7593219,8057119,8424614,7794581}, where Alice and Bob send random pilot sequences to each other and cross-multiply their send and receive signals as the common feature for further secret key generation. The randomness of this common feature, not only involves the random CSI, but is enhanced by Alice's and Bob's random signals, which thereby improves SKR. Despite these advances, the improved SKR are still not enough to meet the current Gbps levels of the transmission rate, which renders as the main challenge to impede PL-SKG from civilian and commercial uses.

To further improve the SKR, reconfigurable intelligent surfaces (RIS) have been studied to increase channel randomness by assigning its configurable phases \cite{9442833,9361290,9360860,staat2020intelligent}. Among these research, the work in \cite{9442833} computes the theoretical SKR of a RIS-secured low-entropy channel, and the works in \cite{9298937,9663196} further design optimal RIS phases set by maximizing the theoretical RIS-involved SKR. 

Despite the RIS's capability to improve PL-SKG to achieve higher SKR, \textcolor{blue}{the study of RIS involving PL-SKG is limited, due to insufficient research specifically focused on malicious RIS.} The threats induced by malicious RISs can be categorized as attackers that aim to destroy or override the reciprocal channel between Alice and Bob. The first category aims to ruin the reciprocal channel-based PL-SKG \cite{9625442}. The second category tries to intercept Alice's and Bob's signals, and reconstruct their common features to crack their secret keys. This physical-layer man-in-the-middle (MITM) RIS eavesdropping was first proposed by \cite{9771319}. Then, the work in \cite{wei2023metasurface} designed RIS hardware to implement the MITM eavesdropping. The work in \cite{10100938} further proposed two MITM-RIS eavesdroppers (MITM-RIS Eves), which, by overriding the legitimate channel with RIS generated channel, can reconstruct the exiting CSI-based and two-way cross-multiplication-based common features, and the secret keys relied upon. 
What is worse, because RIS cannot actively send pilots for protocol \cite{thai2015physical,letafati2020new} and authentication purposes, MITM-RIS attacks are difficult to be detected and countermeasured using standard authentication methods, e.g., hash-based signatures.

To deal with traditional Eves that share part of CSIs with legitimate users, e.g., when the Eve is within the half-wavelength of one legitimate user, the work in \cite{zhou2023physical} developed a domain-adversarial training of an auto-encoder (DAAE). Leveraging the CSIs difference between legitimate users and Eve, DAAE aims to encode Alice's and Bob's reciprocal CSIs into common features that cannot be decoded by a potential Eve. However, in the face of MITM-RIS Eve that can reconstruct most of the legitimate CSIs \cite{10100938}, DAAE will result in a very low available secret key rate (we will show this in Section \ref{compare_daae}). 
\textcolor{blue}{Other similar adversarial learning frameworks, such as generative adversarial networks (GANs), rely on training a discriminator neural network (NN) to distinguish between the output of the generator NN and the real data (e.g., voice, images) that the generator seeks to replicate. This process helps guide the generator to enhance its performance by producing more realistic data. However, the existing GAN framework cannot be directly applied to address the MITM-RIS eavesdropping problem, since the definition of an adversarial discriminator and its corresponding training data in the context of MITM-RIS eavesdropping has not been explored, not to mention obtaining a trustworthy white-box secret key generator for transparent security purposes.}
Nevertheless, these frameworks inspire the design of more sophisticated adversarial learning-based common feature generators to countermeasure the MITM-RIS Eve.

In this work, we exploit the two-way random signals, and design an explainable adversarial learning-based framework to generate legitimate common features that are resistant to the MITM-RIS Eve. The detailed contributions are the following: 
\begin{itemize}
    \item We deduce a theoretical mutual information gap between the legitimate Alice-Bob pair and the MITM-RIS Eve, when Alice and Bob use two-way signals for their common feature generation. This theoretically proves the existence of a common feature space that cannot be reconstructed via the leaked signals to MITM-RIS. 
    \item We design an adversarial learning-based framework for Alice and Bob to learn to derive the common feature space. Here, Alice and Bob are assigned two neural networks (NNs) as feature generators, and an NN of an assumed MITM-RIS Mallory is set as the adversarial part. Cross-correlation is used in the generator's and adversary's loss functions to measure the similarity of features as well as maintain the randomness. The novelty lies in the deployment of adversarial NN, which, by simultaneously training with legitimate generators, can help remove the vulnerable feature spaces that can be learned by MITM-RIS Eves.
    \item To explain the trained common feature generator NNs, we deploy Meijer G function-based symbolic metamodelling \cite{alaa2019demystifying,sun2021scalable} to identify the dominant special terms of dominant neurons. The results then guide us to the designs of explicit formulas-based common feature generators. In this view, we provide a more interpretable and transparent approach for generating physical layer common features and the secret keys they rely on.
    \item We evaluate our proposed adversarial learning-based and explicit formula-based PL-SKG. The results show high key agreement rates between Alice and Bob, and that even the MITM-RIS Eve with the full knowledge of Alice's and Bob's common feature generator (NNs or the explicit formulas), Eve still cannot reconstruct the generated physical layer secret keys. This promising outcome signifies a significant step toward establishing secure wireless communications in the context of untrusted reflective devices, laying the groundwork for future 6G cybersecurity.
\end{itemize}

The rest of this work is structured as follows. In Section II, we describe channel models and the MITM-RIS Eve threats. In Section III, we provide the system overview and elaborate on our design of adversarial learning-based common feature generators. In Section IV, we explain the learned NNs via symbolic meta-modelling, and feature visualization to derive the explicit-formula-based common feature generators. Section V provides a strong NN-based MITM-RIS as the worst-case to test our algorithm. In Section VI, we show our simulation results. We finally conclude this work in Section VII. 

In this work, we use bold lowercase letters for vectors. We use $\|\cdot\|_2$ to denote the $2$-norm, $\text{diag}(\cdot)$ to diagonalize a vector, $\text{det}(\cdot)$ as the determinant of a square matrix, $\circ$ to compute elemental-wise product, and tr$(\cdot)$ to compute the matrix trace. $|\cdot|$ and $(\cdot)^*$ represent the absolute value and the conjugate of a complex value. We denote $[\cdot]_c$ as the modulus operator to take the remainder of a value divided by $c$, and $\lfloor\cdot\rfloor$ to truncate a value. The matrix transpose is denoted as $(\cdot)^T$. $p(\cdot)$, $\mathbb{E}(\cdot)$, and $\mathbb{D}(\cdot)$ represent the probability density function (PDF), the expectation and the variance, respectively. The differential information entropy is denoted as $h(\cdot)$ and the mutual information of two groups of random variables is denoted as $I(\cdot~;\cdot)$. $\Re[\cdot]$ and $\Im[\cdot]$ are to represent the real and imagery parts of a variable. $\mathcal{CN}(\cdot,\cdot)$ is to represent the complex Gaussian distribution with mean and variance.

\section{Problem Formulation}
\begin{figure}[!t]
\centering
\includegraphics[width=3in]{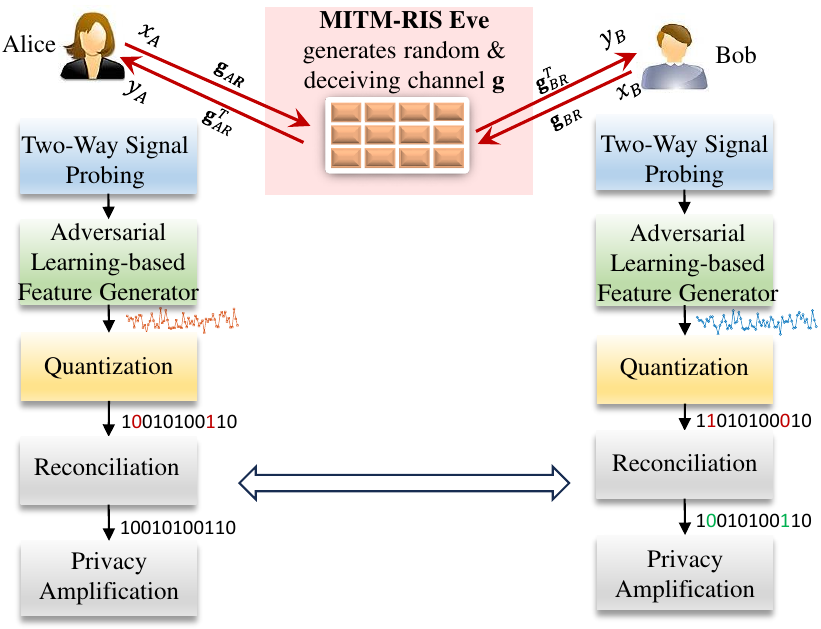}
\caption{Overview of the adversarial learning-based PL-SKG, with MITM-RIS Eve threat. }
\label{fig0}
\end{figure}

In this work, two legitimate users, namely Alice and Bob, \textcolor{blue}{as well as a malicious RIS} are considered, as shown in Fig.~\ref{fig0}.
\textcolor{blue}{Alice and Bob leverage the channel reciprocity and randomness between each other to generate common features for symmetrical secret keys. In the previous work \cite{10100938}, MITM-RIS Eves were designed to insert a deceptive reciprocal and random channel into legitimate Alice-Bob channels, causing Alice and Bob to unknowingly generate physical-layer common features using this MITM-RIS channel, and thereby enabling MITM-RIS Eve to estimate these features and crack the secret keys. To address this, in this work, we consider the worst-case where the Alice-Bob channel is completely composed by malicious RIS, omitting the direct channel between Alice and Bob.} In this way, the malicious RIS, equipped with RF chains \cite{taha2021enabling}, can pursue the MITM attack to reconstruct the legitimate common features and secret keys between Alice and Bob \cite{10100938}. 

\subsection{Channel and Signal Models}\label{system_model}
The legitimate users, Alice and Bob, are considered as single-antenna. The malicious RIS is modelled as a uniform planar array (UPA) of size $M=M_x\times M_y$. The direct channels from Alice and Bob to the RIS, denoted as $\mathbf{g}_{uR}\in\mathbb{C}^{M\times1}~u\in\{A\text{(Alice)},B\text{(Bob)}\}$, are modelled as \cite{9300189}:
\begin{equation}
\label{eq02}
\mathbf{g}_{uR}=\sum_{n=1}^{L}\zeta_{uR,n}\cdot\bm{\xi}\left(\alpha_{uR,n},\beta_{uR,n}\right),
\end{equation}
In (\ref{eq02}), $\zeta_{uR,n}\sim\mathcal{CN}\left(0,C_0\cdot d_{uR}^{-\alpha}\right)$ is the gain for $n$th path between $u$ and the RIS, with $C_0$ the path loss at the reference distance (i.e., $1m$), $d_{uR}$ the line-of-sight (LoS) distance between $u$ to the RIS, and $\alpha$ the exponential loss factor. $L$ is the number of NLoS Rayleigh paths. $\bm{\xi}(\eta,\beta)\triangleq[\exp(j\mathbf{a}(\eta,\beta)\mathbf{l}_1),\cdots,\exp(j\mathbf{a}(\eta,\beta)\mathbf{l}_M)]^T$ is the spatial vector, with $\mathbf{a}(\eta,\beta)\triangleq\frac{2\pi}{\lambda_c}[\sin(\eta)\cos(\beta),\sin(\eta)\sin(\beta),\cos(\eta)]$ and $\mathbf{l}_m\triangleq d\cdot[0,[m-1]_{M_x},\lfloor (m-1)/M_y\rfloor]^T$. $\alpha_{uR,n},\beta_{uR,n}\in[-\pi/2,\pi/2]$ are the half-space elevation and azimuth angles of $n$th path. For the structure of RIS, a square shape element is used with the size as $d\times d$, where $d=\lambda_c/4$ is set (i.e., less than half-wavelength \cite{9300189,8936989}).

With the modelling of the direct channels to RIS, the combined channels from Alice to Bob (Bob to Alice), denoted as $g_{AB}$ ($g_{BA}$), are expressed as:
\begin{equation}
\label{eq1}
\begin{aligned}
    g_{AB}=\mathbf{g}_{BR}^T\cdot \text{diag}(\mathbf{w})\cdot\mathbf{g}_{AR},\\
    g_{BA}=\mathbf{g}_{AR}^T\cdot \text{diag}(\mathbf{w})\cdot\mathbf{g}_{BR}.
\end{aligned}
\end{equation}
In (\ref{eq1}), $\mathbf{w}=\sqrt{A_E}\cdot[\exp(j\varphi_1),\cdots,\exp(j\varphi_M)]^T$ is the phase vector of the RIS, where $\varphi_m\in[0,2\pi)$ with $m\in\{1,\cdots,M\}$ is the phase of $m$th RIS element, and $A_E$ is the amplifying power \cite{taha2021enabling}. 

We then denote Alice's and Bob's sent signals as $x_A,x_B\in\{\sqrt{P_t}\exp(j\iota)|\iota\in[0,2\pi]\}$. According to (\ref{eq1}), their received signals are expressed as:
\begin{equation}
\label{eq3}
\begin{aligned}
y_A=&g_{BA}\cdot x_B+n_A,\\
y_B=&g_{AB}\cdot x_A+n_B,
\end{aligned}
\end{equation}
with $n_A,n_B\in\mathcal{CN}(0,\sigma^2)$ the received noise. To create the worst-case scenario, we assume that all elements of the malicious RIS are equipped with RF chains \cite{taha2021enabling} for receiving signals and base-band signal processing. This extreme setting provides more accurate channel probing and therefore serves as a more powerful RIS Eve than the design in \cite{10100938}, where a sparse RF chain placement and compressed sensing method is used. As such, the RIS received signals from Alice and Bob are expressed as:
\begin{equation}
\label{eve_receive}
\begin{aligned}
\mathbf{y}_R^{(A)}=\mathbf{g}_{AR}\cdot x_A+\mathbf{n}_R^{(A)},\\
\mathbf{y}_R^{(B)}=\mathbf{g}_{BR}\cdot x_B+\mathbf{n}_R^{(B)},
\end{aligned}
\end{equation}
where $\mathbf{n}_R^{(A)},\mathbf{n}_R^{(B)}\in\mathcal{CN}(0,\sigma^2\mathbf{I}_M)$ are the received noises.

\subsection{MITM-RIS Eve Threats to Existing PL-SKGs}
In a standard PL-SKG framework, Alice and Bob first carry out channel probing, i.e., they leverage their reciprocal channel properties to individually extract common features, which will then be quantitized to generate symmetrical secret keys. There are two common channel probing approaches:
\begin{itemize}
    \item CSI-based Common Features.
    \item Two-Way Cross Multiplication-based Common Features.
\end{itemize}
In the previous work \cite{10100938}, two MITM-RIS Eve schemes are proposed to compromise these two approaches, which will be elaborated below.

{\color{blue}
\begin{Rem}
The core idea of MITM-RIS Eve \cite{10100938} is to insert a deceptive reciprocal and random channel into the legitimate Alice-Bob link by reflection. By increasing its reflection power, i.e., $A_E$, MITM-RIS Eve can make the Alice-RIS-Bob path dominate over the direct Alice-Bob channel. This manipulation leads Alice and Bob to unknowingly use this overridden channel for common feature generation, allowing MITM-RIS Eve (equipped with RF chains for receiving) to estimate these features and compromise the secret keys.
\end{Rem}}

{\color{blue}
\begin{Rem}
Different from untrusted relays, MITM-RIS Eve \cite{10100938}, given its reflective property (unable to actively send pilots), is difficult to be detected and addressed using the standard authentication methods, e.g., hash-based signatures, and the relay pilot-based countermeasures in \cite{thai2015physical,letafati2020new}.
\end{Rem}}

\subsubsection{MITM-RIS Eve against CSI-based Common Features}
In CSI-based PL-SKG, Alice and Bob send pre-defined pilot sequences and estimate the reciprocal CSI in two consecutive time slots that maintain channel reciprocity, i.e., $g=g_{AB}=g_{BA}$ \cite{9442833,9361290,9360860,staat2020intelligent}. {\color{blue}The estimated CSIs based common feature at Alice and Bob are \cite{9442833,9361290,9360860,staat2020intelligent}:
\begin{equation}
\label{legitimate1}
\begin{aligned}
    \hat{h}_A&=y_A\cdot x_B^*/P_t=g_{BA}+x_B^*/P_t\cdot n_A,\\
    \hat{h}_B&=y_B\cdot x_A^*/P_t=g_{AB}+x_A^*/P_t\cdot n_B.
\end{aligned}
\end{equation}
The reciprocal CSIs, i.e., $\hat{h}_A\approx\hat{h}_B$, can be leveraged as the common features to further generate symmetrical secret keys.}

{\color{blue}
\begin{Rem}
For the CSI based common feature generation, MITM-RIS Eve \cite{10100938} can recover the legitimate Alice's and Bob's reciprocal CSIs (i.e., $g=g_{AB}=g_{BA}$) via her received signals from Alice and Bob $\mathbf{y}_R^{(A)}$ and $\mathbf{y}_R^{(B)}$, i.e., 
\begin{equation}
\label{eve-csi}
\begin{aligned}
\hat{\mathbf{g}}_{AR}&=\mathbf{y}_R^{(A)}\cdot x_A^*/P_t=\mathbf{g}_{AR}+x_A^*/P_t\cdot\mathbf{n}_R^{(A)},\\
\hat{\mathbf{g}}_{BR}&=\mathbf{y}_R^{(B)}\cdot x_B^*/P_t=\mathbf{g}_{BR}+x_B^*/P_t\cdot\mathbf{n}_R^{(B)},\\
\hat{h}_E&=\hat{\mathbf{g}}_{BR}^T\cdot\text{diag}(\mathbf{w})\cdot\hat{\mathbf{g}}_{AR}= g+\varepsilon_E,
\end{aligned}
\end{equation}
with $\varepsilon_E=[x_A^*\mathbf{g}_{AR}^T \text{diag}(\mathbf{w})\mathbf{n}_{R}^{(B)}+x_B^*\mathbf{g}_{BR}^T \text{diag}(\mathbf{w})\mathbf{n}_{R}^{(A)}]/P_t+x_A^*x_B^*/P_t^2(\mathbf{n}_{R}^{(A)})^T \text{diag}(\mathbf{w})\mathbf{n}_{R}^{(B)}$ the noise component. The results in  \cite{10100938} show a $90\%$ match rate of MITM-RIS Eve reconstructed secret keys with legitimate secret keys, when MITM-RIS uses $30$dB reflection power. This shows the secret keys using the reciprocal CSI-based common features between Alice and Bob are unsecured. 
\end{Rem}}

\subsubsection{MITM-RIS Eve against Two-Way Cross-Multiplication-based Common Features}\label{two-way}
In the two-way cross-multiplication PL-SKG, Alice and Bob send random sequences in two consecutive time slots that maintain channel reciprocity, and multiply their send and receive signals as common features. {\color{blue}The two-way cross-multiplication-based common features generated at Alice and Bob are \cite{8424614,8798662}:
\begin{equation}
\label{cross_multi}
\begin{aligned}
\phi_A= x_A\cdot y_A= g\cdot x_B\cdot x_A+x_A\cdot n_A,\\
\phi_B=x_B\cdot y_B= g\cdot x_B\cdot x_A+x_B\cdot n_B.
\end{aligned}
\end{equation}
Then, secret keys can be generated at Alice and Bob via their common features $\phi_A\approx\phi_B$. }

{\color{blue}
\begin{Rem}
For the two-way cross-multiplication-based common feature generation, MITM-RIS Eve \cite{10100938} can reconstruct the legitimate Alice's and Bob's common feature by directly multiplying the received signals from Alice and Bob, i.e., 
\begin{equation}
\phi_E=\left(\mathbf{y}_R^{(B)}\right)^T\cdot \text{diag}(\mathbf{w})\cdot\mathbf{y}_R^{(A)}= g\cdot x_B\cdot x_A+\epsilon_E,
\end{equation}
where $\epsilon_E=x_A\mathbf{g}_{AR}^T \text{diag}(\mathbf{w})\mathbf{n}_{R}^{(B)}+x_B\mathbf{g}_{BR}^T \text{diag}(\mathbf{w})\mathbf{n}_{R}^{(A)}+(\mathbf{n}_{R}^{(A)})^T \text{diag}(\mathbf{w})\mathbf{n}_{R}^{(B)}$ is the noise component. The results in   \cite{10100938} show a $90\%$ match rate of MITM-RIS Eve reconstructed secret keys with legitimate secret keys, when MITM-RIS uses $30$dB reflection power. This demonstrates the insecure property of the two-way cross-multiplication-based common features between Alice and Bob.
\end{Rem}}

\section{Adversarial Learning-based Common Feature Generator Design}\label{gann}
In order to address the above-mentioned challenges, in this paper, we design the adversarial learning-based PL-SKG against the MITM-RIS Eve. As is shown in Fig.~\ref{fig0}, the framework contains (i) the two-way signal probing, (ii) adversarial learning-based common feature generator, (iii) quantization \cite{6171198,mathur2008radio}, (iv) reconciliation \cite{10.1007/3-540-48285-7_35}, and (v) privacy amplification \cite{impagliazzo1989pseudo}. In this work, we focus on the adversarial learning-based common feature generator, which converts Alice and Bob two-way random signals into common features that cannot be learned by the MITM-RIS Eve. This step serves as the seed for the following standard secret key generation steps (iii)-(v).

Similar to the two-way cross-multiplication method described in Section \ref{two-way}, the probing signals are the two-way random signals, i.e., $x_A$ and $x_B$, sent by Alice and Bob in two consecutive time slots that maintain channel reciprocity. To be specific, in the first time-slot, Alice sends random $x_A$ and Bob receives $y_B$. Then in the second time-slot, Bob sends random $x_B$ and Alice receives $y_A$. In the meantime, MITM-RIS Eve can receive the signals from Alice and Bob as $\mathbf{y}_R^{(A)}$ and $\mathbf{y}_R^{(B)}$, respectively in the two time-slots. The expressions of $y_A$, $y_B$, $\mathbf{y}_R^{(A)}$ and $\mathbf{y}_R^{(B)}$ are provided in (\ref{eq3})-(\ref{eve_receive}). Next, Alice and Bob will individually use their two-way signals $(x_A,y_A)$ and $(x_B,y_B)$ to generate common features that cannot be learned by MITM-RIS Eve via $\mathbf{y}_R^{(A)}$ and $\mathbf{y}_R^{(B)}$. 

\subsection{Theoretical Positive SKR of Two-way Signals}
The rate of secret key generation, i.e., SKR, is defined by the mutual information difference between Alice's and Bob's two-way signals, and MITM-RIS Eve's received signals. The existence of the common feature space that cannot be learned by MITM-Eve can be proved by the positive value of SKR, i.e., 
\begin{equation}
\label{theo1}
\begin{aligned}
\text{SKR}&=I\left(
x_A,y_A;
x_B,y_B
\right)-\underset{a\in\{A,B\}}{\text{max}}I\!\left(
x_a,y_a;
\mathbf{y}_R^{(A)},\mathbf{y}_R^{(B)},\mathbf{w}\!
\right)\\
&>0. 
\end{aligned}
\end{equation}
where the detailed proof is provided in Appendix~\ref{appendix1}. \textcolor{blue}{Note that this secret key rate analysis in (\ref{theo1}) and Appendix~\ref{appendix1} is not for the specific two-way cross-multiplication-based common features (i.e., one's sent signal multiplying its received signal). Instead, it proves the existence of the mutual information gap for the general two-way signals that can be used by Alice and Bob for common feature generation. This underpins the following adversarial learning-based common feature generator designs, and the existence of common feature space that cannot be learned by MITM-RIS Eve.}

\begin{figure*}[!t]
\centering
\includegraphics[width=7in]{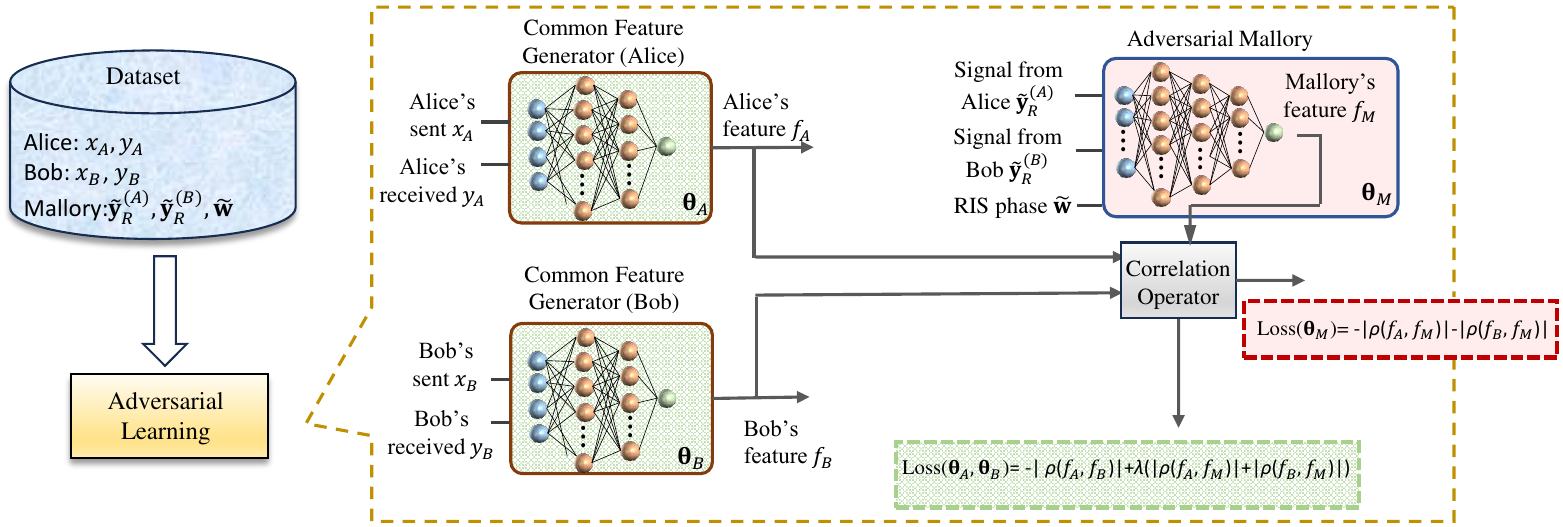}
\caption{The designed adversarial learning structure to train Alice's and Bob's common feature generators, including common feature generator NNs and adversarial NN (an assumed MITM-RIS Mallory by Alice and Bob). }
\label{fig1}
\end{figure*}

\subsection{Structure of Adversarial Learning-based Design}
Leveraging the theory showing the existence of the common feature space that is resistant to MITM Eve, the adversarial learning-based framework is established. The approach includes Alice's and Bob's common feature generators.
We also deliberately created an adversarial MITM-RIS Mallory that is designed to cover all the spatial positions between Alice and Bob. Note that the adversarial Mallory is not the real MITM-RIS Eve. Whilst in practice the MITM-RIS Eve will not be cooperative, we can emulate different statuses of the Eve in an offline manner during the training stage. 
This allows us to derive a common feature space of Alice and Bob that cannot be learned by \textcolor{blue}{any MITM-RIS that is spatially located between Alice and Bob.}

\subsubsection{Legitimate Feature Generators of Alice and Bob}
\textcolor{blue}{The legitimate common feature generators of Alice and Bob are defined as two fully connected NNs, denoted as $\Psi_{\bm{\theta}_A}(\cdot)$ and $\Psi_{\bm{\theta}_B}(\cdot)$ with parameters $\bm{\theta}_A$ and $\bm{\theta}_B$, respectively.} The two NNs have the same structures. 
The feature generator NNs of Alice and Bob are expressed as
\begin{equation}
\label{eq7}
\begin{aligned}
f_A&=\Psi_{\bm{\theta_A}}\Big(\Re[x_A], \Im[x_A], \Re[y_A], \Im[y_A]\Big),\\
f_B&=\Psi_{\bm{\theta_B}}\Big(\Re[x_B], \Im[x_B], \Re[y_B], \Im[y_B]\Big),
\end{aligned}
\end{equation}
where $f_A,f_B$ are the common features of Alice and Bob, respectively.   

The structure of the feature generator NN is explained as follows.
\begin{itemize}
    \item Input layer: 4 neurons, which are the real and imaginary of the user's send and receive signals, i.e., $\Re[x_u], \Im[x_u], \Re[y_u], \Im[y_u]$, with $u\in\{A,B\}$.
    \item Hidden layers: two layers with $512$ and $128$ neurons, respectively. The activation function is ReLU.
    \item Output layer: 1 neuron, activated by Sine function, and output $f_u$.
\end{itemize}

\subsubsection{Adversarial Network of Mallory}
To prevent MITM-RIS from reconstructing the legitimate features, Alice and Bob further assume an adversary, the MITM-RIS Mallory. By denoting the parameters of the adversarial NN as $\bm{\theta}_M$, its output is expressed as:
\begin{equation}
\label{eq8}
\begin{aligned}
f_M=\Psi_{\bm{\theta}_M}\Big(\Re\left[\tilde{\mathbf{y}}_{R}^{(A)}\right], &\Im\left[\tilde{\mathbf{y}}_{R}^{(A)}\right], \Re\left[\tilde{\mathbf{y}}_{R}^{(B)}\right], \\
&\Im\left[\tilde{\mathbf{y}}_{R}^{(B)}\right], \Re[\tilde{\mathbf{w}}], \Im[\tilde{\mathbf{w}}]\Big),
\end{aligned}
\end{equation}
where $\tilde{\mathbf{y}}_{R}^{(A)}=\tilde{\mathbf{g}}_{AR}x_A+\tilde{\mathbf{n}}_R^{(A)}$ and $\tilde{\mathbf{y}}_{R}^{(B)}=\tilde{\mathbf{g}}_{BR}x_B+\tilde{\mathbf{n}}_R^{(B)}$ are denoted as the received two-way signals at the assumed Mallory RIS (with $\tilde{\mathbf{g}}_{AR}$$\tilde{\mathbf{g}}_{BR}$ the channels from Alice and Bob to the assumed Mallory RIS), and $\tilde{\mathbf{w}}$ is Mallory RIS phase, in order to distinguish the variables related to real Eve. 

The structure of the adversarial NN is given below.
\begin{itemize}
    \item Input layer: $6M$. The inputs are the simulated RIS's phase (real and imaginary parts) and received signals from Alice and Bob, i.e., $\tilde{\mathbf{g}}_{AR}\cdot x_A$ and $\tilde{\mathbf{g}}_{BR}\cdot x_B$. The input terms used here represent the information leakage from Alice and Bob that can be exploited by MITM-RIS to obtain the legitimate common features.
    \item Hidden layer: three layers with $1024$, $512$, and $128$ neurons, respectively, with ReLU activation function. 
    \item Output layer: 1 neuron activated by Sine function, and output $f_M$.
\end{itemize}

\textcolor{blue}{It is noteworthy that the MITM-RIS Mallory is different from the real MITM-RIS Eve. In general adversarial learning framework, the objective is to train legitimate NNs that can resist specific adversarial attacks. To achieve this, legitimate parties often simulate such attacks through adversarial NNs that will be trained simultaneously with the legitimate NNs, like what we do by emulating MITM-RIS Mallory. In our work, the MITM-RIS Mallory represents the intentionally emulated man-in-the-middle RIS eavesdropper, deliberately designed by the legitimate parties to serve as a model adversary. By incorporating Mallory into the legitimate training process, the legitimate common feature generator NNs can be trained to learn a common feature space of Alice and Bob that cannot be learned by any real MITM-RIS Eves (which will be tested in the Simulation section). }

\subsection{Dataset Construction}\label{train_data}
Alice, Bob, and the assumed Mallory RIS are modelled according to the description in Section \ref{system_model}. From (\ref{eq02}), the distribution of channels between one legitimate user to the RIS is determined by their LoS distance and the elevation and azimuth angles. In this view, to encompass comprehensive scenarios, we evenly split (i) the angle half-space $[-\pi/2, \pi/2]$ by $100$, and (ii) the distance from $0$m to $200$m by $1000$. Then, a total of $N=10^7$ pairs of $(\tilde{\mathbf{y}}_R^{(A)}, \tilde{\mathbf{y}}_R^{(B)}, \tilde{\mathbf{w}}, x_A, x_B, y_A, y_B)$ is sampled as the dataset derived from the channel and signal models in (\ref{eq02})-(\ref{eq3}). In the dataset, $70\%$ is used for training, $20\%$ is for validation, and the rest $10\%$ is for testing. 

\subsection{Loss Function}
The loss functions of generator NNs and the adversarial NN have opposite objectives. The loss function of the adversarial Mallory NN is:
\begin{equation}
\label{eq9}
\begin{aligned}
\text{Loss}(\bm{\theta}_M)= -|\rho(f_A, f_M)| - |\rho(f_B, f_M)|,
\end{aligned}
\end{equation}
where the correlation coefficient is expressed by: 
\begin{equation}
\label{corr_coef}
\rho(X, Y)=\frac{\mathbb{E}(XY)-\mathbb{E}(X)\cdot\mathbb{E}(Y)}{\sqrt{\mathbb{D}(X)\cdot\mathbb{D}(Y)}}.
\end{equation}
The aim is to let adversarial Mallory learn the legitimate features, so that the vulnerable features can be further discarded by the training of legitimate generators. The loss function here is designed by maximizing the absolute correlation coefficients between Alice's (Bob's) and adversarial Mallory's features in one batch of training data. 

The loss function of Alice and Bob contains two objectives (i) to maximize the absolute correlation coefficient between Alice's and Bob's features among one batch of training data, and (ii) to ensure the adversarial Mallory cannot reconstruct the legitimate features from their received signals, i.e.,
\begin{equation}
\label{eq10}
\begin{aligned}
\text{Loss}(\bm{\theta}_A,\bm{\theta}_B)=& - |\rho(f_A, f_B)|
\\&-\lambda \left[-|\rho(f_A, f_M)| - |\rho(f_B, f_M)|\right].
\end{aligned}
\end{equation}
where $\lambda=0.8$ is the assigned trade-off coefficient (see Appendix \ref{trade_off} for other values). In (\ref{eq10}), $f_M$ from the Mallory NN serves as the adversarial part. By minimizing the correlations with the assumed adversarial Mallory, i.e., $|\rho(f_A, f_M)|+|\rho(f_B, f_M)|$, legitimate feature generator NNs can avoid constructing features that can be learned by potential real Eves.

It is noteworthy that, in these loss functions, we characterize the commonality of features via correlation coefficients other than the basic mean squared error (MSE). The reasons are two-fold. First, MSE only accounts for the difference between two features, but is unable to characterize the mutual information. For example, if Alice's and Bob's features are with all $1$ elements, their MSE will approach $0$ but their features lack randomness and mutual information. In contrast, the correlation coefficient in (\ref{corr_coef}) involves not only the similarity (in the numerator) but also the variances (in the denominator) to consider the randomness, which thereby serves as the better way to characterize the (first-order) mutual information. Secondly, achieving the MSE criterion is possible by directly scaling down the outputs of $f_A$ and $f_B$. However, it's important to note that whilst the MSE may decrease, e.g., $\text{MSE}(0.1\cdot f_A,0.1\cdot f_B)<\text{MSE}(f_A,f_B)$, the mutual information remains unchanged, i.e., $I(0.1\cdot f_A;0.1\cdot f_B)=I(f_A;f_B)$. \textcolor{blue}{The performance of MSE based loss function is provided in Appendix \ref{appendixC}.}

\subsection{Algorithm Flow of Training Process}
In the training stage, the batch size is set as $B=64$ and the Adam optimizer is utilized with a learning rate of $10^{-5}$.

The overall training process is detailed in Algorithm \ref{algo1}. Given the number of training data, learning rate, maximum epoch number and batch size, step 1 is to generate the training data for Alice's and Bob's generator NNs and the adversarial Mallory NN. Step 2 is to initialize these $3$ NNs. The training starts from steps 3 to 11, by sampling a batch of training data in step 4, computing the outputs of NNs in steps 5-7, and updating the adversarial NN (step 8) and the generator NNs (step 9) respectively. It is noteworthy that, in the updating stage of Alice's and Bob's NNs, adversarial Mallory's model is detached and not updated, and vice versa. The outputs are the trained Alice's and Bob's feature generator NNs which can (i) generate common features from the two-way signals, and (ii) prevent the MITM-RIS Eve from reconstructing the common features via the RIS's received signals (we will further test this in Section IV and Simulations). 
\begin{algorithm}[t]
\label{algo1}
\caption{Training of Adversarial Feature Construction NNs}
\LinesNumbered
\KwIn {number of Training data $N$, learning rate $r=10^{-5}$, episode number $E_{\text{max}}=10^4$, batch size $B=64$}
\KwOut{Alice's and Bob's feature generator NNs $\Psi_{\bm{\theta}_A}(\cdot)$ and $\Psi_{\bm{\theta}_B}(\cdot)$.}

Training data: Sampling a total of $N$ tuples of $(\tilde{\mathbf{y}}_R^{(A)}, \tilde{\mathbf{y}}_R^{(B)}, \tilde{\mathbf{w}}, g_{AB}, x_A, x_B, y_A, y_B)$ according to (\ref{eq02})-(\ref{eq3}), where the elevation and azimuth angles and LoS distances are evenly selected given the split grids of angle and distance spaces\;

Initialize Alice's and Bob's feature generator NNs and adversarial Mallory NN as $\Psi_{\bm{\theta}_A}(\cdot)$, $\Psi_{\bm{\theta}_B}(\cdot)$, $\Psi_{\bm{\theta}_M}(\cdot)$\;

\For{epoch $=1,\cdots,E_{\text{max}}$}{
Sampling a batch $B=128$ of training data \;
\For{each $i=1,\cdots,B$ batch}{
Compute $f_A[i]$ and $f_B[i]$ via (\ref{eq7})\;
Compute $f_M[i]$ via (\ref{eq8})\;
Update adversarial Mallory NN by $\text{Loss}(\bm{\theta}_M)$ in (\ref{eq9})\;
Update Alice's and Bob's feature generator NNs by $\text{Loss}(\bm{\theta}_A,\bm{\theta}_B)$ in (\ref{eq10})\;
}
}
\end{algorithm}

{\color{blue}
\subsection{Computational Complexity Analysis}
The computational complexity consists of two parts: (i) the training process and (ii) the operation of common feature generators after training. The training process is a one-time procedure, so its complexity does not impact the application of common feature generation for Alice and Bob once training is complete. For the operational phase of common feature generation, the complexity depends on the structure of the feature generator NN. Based on the structure detailed in Section IV.A, the number of multiplication operations required is $4\times512+512\times128+128=67,712$, which is feasible with 1GB RAM edge devices \cite{murshed2021machine}.}

\section{Explainability to Explicit Common Feature Generator Formula}
The opaque black-box nature of the trained common feature generator NNs is unable to present trustworthiness for their practical usages. The explainability is even more important for key generation as it is a network security mechanism.

\textcolor{blue}{In this section, we present the process to interpret the learned common feature generator NNs and further to design an explicit formula-based feature generator leveraging these insights. To do so, we first leverage the symbolic representation of Meijer G-function \cite{alaa2019demystifying,sun2021scalable} to represent the selected important neurons. Next, we visualize the learned feature space, allowing us to deduce a concise and explicit formula for the common feature generator. This formula acts as a white-box solution to defend against MITM-RIS eavesdropping.}

\subsection{Symbolic Representation via Meijer G Functions}
\subsubsection{Introduction of Meijer G-function} 
In essence, the Meijer G-function belongs to a family of univariate functions, each of which corresponds to a linear combination of certain special functions. The equation of the Meijer G function is given as \cite{alaa2019demystifying}
\begin{equation}
\begin{aligned}
&G_{p,q}^{m,n}\left(\substack{a_1,\cdots,a_p\\b_1,\cdots,b_q}|x\right)\\
=&\frac{1}{2\pi j}\int_{\mathcal{L}}\frac{\prod_{i=1}^{m}\Gamma(b_i-s)\prod_{i=1}^{n}\Gamma(1-a_i+s)}{\prod_{i=m+1}^{q}\Gamma(1-b_i+s)\prod_{i=n+1}^{p}\Gamma(a_i-s)}(x)^sds,
\end{aligned}
\end{equation}
where $0\leq m\leq q,0\leq n \leq p$ are predefined integer numbers, $a_1\cdots a_p$ and $b_1,\cdots,b_q$ are continuous real parameters with relations $a_k-b_i\notin \mathbb{Z}^+$ for $k=1,\cdots,n$ and $i=1,\cdots,m$ and $x\neq 0$. Here $\Gamma(\cdot)$ is the Gamma function extended on complex variable $s$. The integral path $\mathcal{L}$ is from $-j\infty$ to $j\infty$ separate the poles of the factors $b_i-s$ from those of factors $1-a_k+s$. Some of the equivalent explicit forms of Meijer G functions when specifying the parameters $m,n,p,q$, $a_1,\cdots,a_p$, and $b_1,\cdots,b_q$ are \cite{alaa2019demystifying}
\begin{equation}
\label{eq17}
\begin{aligned}
&G_{3,1}^{0,1}\left(\substack{2, 2, 2\\1}|x\right)\equiv x\\
&G_{0,1}^{1,0}\left(\substack{-\\0}| x\right)\equiv \exp(x)\\
&G_{2,2}^{1,2}\left(\substack{1, 1\\1, 0}|x\right)\equiv \log(1+x). 
\end{aligned}
\end{equation}

\subsubsection{Meijer G-based Representation to Fit a Neuron}
\textcolor{blue}{Kolmogorov superposition theorem states that every multivariate continuous function can be written as a finite two-layer composition of univariate continuous functions. This helps avoid using too complicated composite Meijer G functions to fit the trained NNs. As such, we adopt the two-layer Meijer G function-based representation framework in \cite{alaa2019demystifying}}, i.e.,
\begin{equation}
\label{mG}
G(\mathbf{x}, \bm{\theta}_G)=\sum_{l=0}^1 g_{1,l}\left(\bm{\theta}_{1,l}\Big|\sum_{i=0}^3 g_{0,i}(\bm{\theta}_{0,i}|x_i)\right),
\end{equation}
where $\mathbf{x}\!=\![x_1,x_2,x_3,x_4]\!\!=\!\![\Re\{x_A\}, \Im\{x_A\}, \Re\{y_A\}, \Im\{y_A\}]$, which is the input of Alice's feature generator NN.
In (\ref{mG}), the function has $2$ layers, where the first layer consists of $4$ univariate Meijer G functions, i.e., $g_{0, i}(\bm{\theta}_{0,i}|x_i)\triangleq G_{2,2}^{1,2}(\bm{\theta}_{0,i}|x_i)$, and the second layers are $2$ univariate Meijer G functions, i.e., $g_{1, l}(\bm{\theta}_{1,l}|\cdot)\triangleq G_{0,1}^{1,0}(\bm{\theta}_{1,l}|\cdot)$ with the inputs as the summation of first layer outputs. Here, $\bm{\theta}_G\triangleq[\bm{\theta}_{1, 0},\bm{\theta}_{1, 1},\bm{\theta}_{0, 0},\bm{\theta}_{0, 1},\bm{\theta}_{0, 2},\bm{\theta}_{0, 3}]$ are the parameters that use to fit the neurons of Alice's feature generator NN. 

As we denote the output of the $n$th neuron at $m$th layer of the trained Alice's generator NN as $\Psi_{\bm{\theta}_A}^{(m,n)}(\mathbf{x})$, 
the loss function to fit it is expressed as:
\begin{equation}
\label{lossG}
\text{Loss}_G(\bm{\theta}_G) = \left\|G(\mathbf{x}, \bm{\theta}_G)-\Psi_{\bm{\theta}_A}^{(m,n)}(\mathbf{x})\right\|_2^2.
\end{equation}
Then, the Adam optimizer is adopted to minimize the above loss function and derive $\bm{\theta}_G$. With the obtained $\bm{\theta}_G$, $G(\mathbf{x},\bm{\theta}_G)$ can be then expressed as the summation of special terms of Meijer G functions such as the terms in (\ref{eq17}).

\subsubsection{Implementation}
\textcolor{blue}{Given the structural complexity of the trained common feature generator NNs, we avoid directly representing their complicated final output. Instead, we implement the Meijer G representation to fit the important neurons of the trained Alice's feature generator NNs, and analyze the occurrence of special Meijer G terms, which then informs the design of an explicit formula-based common feature generator.}

\textbf{Step 1:} We run the trained Alice's feature generator NN via all the data in the validation dataset. \textcolor{blue}{We select all neurons from the hidden layers, whose accumulated activated values are larger than $0$.} 

\textbf{Step 2:} Each selected neuron is fitted using the representative framework in (\ref{mG}) and the loss function in (\ref{lossG}). With the obtained parameters $\bm{\theta}_G$, the fitted $G(\mathbf{x},\bm{\theta}_G)$ is expressed in terms of the explicit terms of equivalent Meijer G-functions. 

\textbf{Step 3:} We count the appearances of special terms from the obtained parameters in step 2, for all the selected neurons. 

\begin{figure}[!t]
\centering
\includegraphics[width=3.4in]{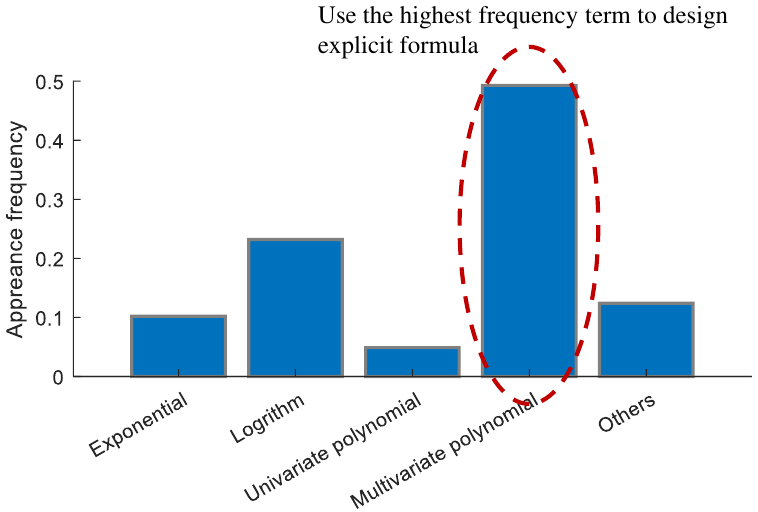}
\caption{The appearance frequency of the special terms derived from fitted Meijer G functions.}
\label{fig:xAI_func}
\end{figure}

The appearance frequency of special terms is plotted in Fig.~\ref{fig:xAI_func}.  It is seen that the percentage of the polynomial terms is obviously higher than others. 
\textcolor{blue}{It is true that the complexity of the Meijer G-based representation framework affects the interpretation of the trained neurons. However, as we identify the dominant Meijer G terms as polynomial terms, we will only leverage the polynomial form to design the explicit formula-based common feature generator. In this sense, the complexity of the Meijer G symbolic metamodeling will not directly influence the final explicit formulas.}

{\color{blue}
\subsection{Insights from Learned Common Features}

\begin{figure}[!t]
\centering
\includegraphics[width=3.4in]{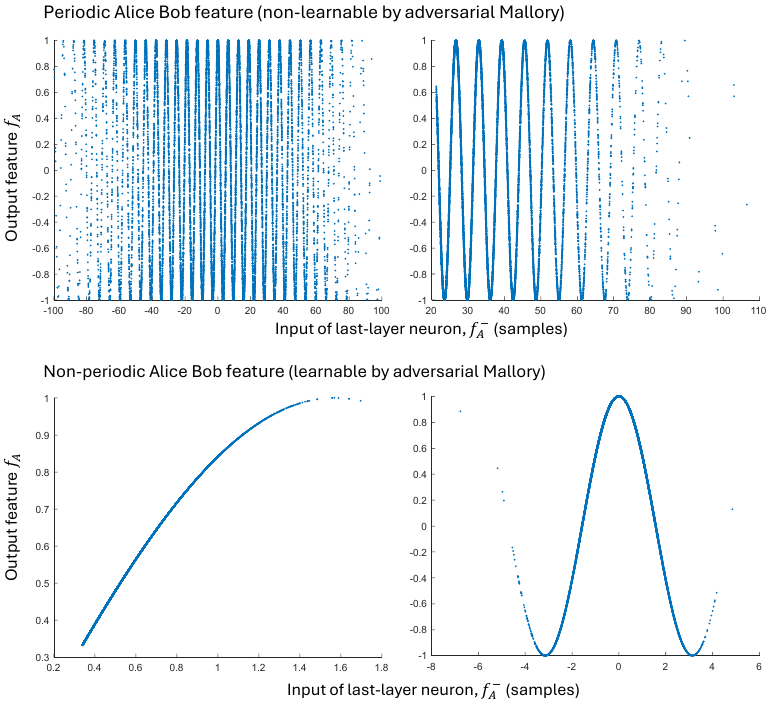}
\caption{Visualization of common features generated by trained Alice and Bob generator NNs: periodic features that are non-learnable by adversarial Mallory, and non-periodic features that are learnable by adversarial Mallory.}
\label{feature_space}
\end{figure}

We next visualize the common feature spaces learned by Alice and Bob, which can and cannot be learned by the assumed adversarial Mallory. In Fig. \ref{feature_space}, the x-axis represents the input points of the last (output) layer neuron, denoted as $f_A^-$ for Alice and $f_B^-$ for Bob, and the y-axis shows the corresponding output features, i.e., $f_A$ and $f_B$. From Fig. \ref{feature_space}, the secured Alice and Bob common features that cannot be learned by MITM-RIS Mallory are those exhibiting periodic properties. In contrast, Mallory can learn to reach out to the features that lack periodicity (e.g., in Fig. \ref{feature_space} the features with no or only one segment of periodicity). The reason is explained in the following two-fold. 

\subsubsection{Polynomial-based Feature Bettering}
Before going into the Sine activation function of the output-layer neuron, adversarial Mallory can learn $f_M^-$, which is highly correlated with $f_A^-$ and $f_B^-$. Despite that, the trained legitimate common feature generator NNs can approach an enlarged correlation coefficient gap, compared with the two-way cross-multiplication based feature ($\phi_A$ $\phi_B$ and $\phi_E$ in Section \ref{two-way}), i.e., 
\begin{equation}
\label{enlarged_cc}
\begin{aligned}
&|\rho(f_A^-,f_B^-)|-\max\{|\rho(f_A,f_M^-)|,|\rho(f_B,f_M^-)|\}\\
&>|\rho(\phi_A,\phi_B^-)|-\max\{|\rho(\phi_A,\phi_E)|,|\rho(\phi_B,\phi_E)|\},
\end{aligned}
\end{equation}
Such learned features $f_A^-$ and $f_B^-$ will then go into the Sine activation function of the output neuron to further cancel their correlation with MITM-RIS Mallory, which will be explained below. 

\subsubsection{Periodical Function-based Feature Bettering}
The periodical function can help cancel the correlation of legitimate common features with adversarial MITM-RIS. As is shown in Fig. \ref{feature_space}, one example of obtained features before the last-layer neuron is $$f_A^-=[1.7785,-11.5793,-15.6173,30.9349,-6.5427],
$$ $$f_B^-=[1.6977,-12.0710,-15.3097,30.7069,-7.0318],$$ 
$$f_M^-=[1.1545,-13.5137,-17.6408,28.7752,-6.3973],$$ with $|\rho(f_A^-,f_B^-)|=0.9998$, $|\rho(f_A^-,f_M^-)|=0.9985$, and $|\rho(f_B^-,f_M^-)|=0.998$. Then, we notice that $$[f_A^-]_{2\pi}=[1.7785,0.9871,3.2323,5.8021,6.0237],$$ $$[f_B^-]_{2\pi}=[1.6977,0.4953,3.5398,5.5741,5.5345
],$$ $$[f_M^-]_{2\pi}=[1.1545,5.3359,1.2087,3.6424,6.1691]$$ with $|\rho([f_A^-]_{2\pi},[f_B^-]_{2\pi})|=0.989$, $|\rho([f_A^-]_{2\pi},[f_M^-]_{2\pi})|=0.336$ and $|\rho([f_B^-]_{2\pi},[f_M^-]_{2\pi})|=0.208$. By analyzing the data, we uncover the fact that the correlation coefficient between legitimate Alice (Bob) and adversarial Mallory can be canceled by omitting their common quotients that share large amounts of similarity.

Further, let $f_u^-=\overline{f}_u^-+[f_u^-]_{2\pi},~u\in\{A,B,M\}$, with modulus $2\pi$. The sine activation function at the output neuron operates:  
\begin{equation}
\label{sin}
\begin{aligned}
f_u=&\sin(f_u^-)=\sin\left(\overline{f}_u^-+[f_u^-]_{2\pi}\right)=\sin(2n\pi+[f_u^-]_{2\pi})\\
=&\sin([f_u^-]_{2\pi}).
\end{aligned}
\end{equation}
From the last term of (\ref{sin}), sine activation function can be treated as a pathway to implement the modulo process, with the outputs:
$$f_A=[0.9785,0.8344,-0.0906,-0.4627,-0.2566],$$
$$f_B=[0.9920,0.4753,-0.3878,-0.6511,-0.6807],$$
$$f_M=[0.9146,-0.8119,0.9352,-0.4801,-0.1138],$$
and further enlarged correlation coefficient gap, i.e., $|\rho(f_A,f_B)|=0.978$, $|\rho(f_A,f_M)|=0.139$ and $|\rho(f_B,f_M)|=0.238$. 

Leveraging the insights of (i) the polynomial dominant feature terms, and (ii) the periodic function to exclude signal correlations with adversarial MITM-RIS, we will design the explicit-formula-based common feature generator in the next part.}

\subsection{Explicit Formula of the Common Feature Generator}
We highlight the insights from the learned common features space: The secured Alice and Bob common features are obtained by (i) polynomial based construction with important neurons represented by polynomial Meijer-G terms, and (ii) the periodical activation function to cancel the correlation between legitimate Alice (Bob) and adversarial Mallory. Leveraging these, we formulate explicit legitimate Alice and Bob common feature generator through the following $3$ steps. 

{\color{blue}
First, we denote the polynomial base (as dominated in Meijer G-based symbolic representation) for Alice and Bob (i.e., $u\in\{A,B\}$ as:
\begin{align}
\label{mm1}
\mathbf{z}_u=&\left[\Re\{x_u\}^m\cdot\Im\{x_u\}^n\cdot\Re\{y_u\}^p\cdot\Im\{y_u\}^q\right]^T,\nonumber\\
&~m,n,p,q\in\{0,1,2\}
\end{align}
Here, we omit the polynomial terms with ($m,n,p,q\geq3$), given that the magnitudes of send and receive signals are less than $1$, i.e., $|x_A|,|y_A|<1$. 

Second, we use the linear combination of $\mathbf{z}_u$ to fit Alice's and Bob's generated features before the output neuron, i.e., $f_u^-$. The least square is used to obtain the linear coefficients $\bm{\varrho}$, i.e., 
\begin{equation}
\bm{\varrho}=\argmin\left\|f_u^--\bm{\varrho}^T\cdot\mathbf{z}_u\right\|_2^2. 
\end{equation}

Third, the fitted $f_A^-$ and $f_B^-$ by $\bm{\varrho}^T\cdot\mathbf{z}_A$ and  $\bm{\varrho}^T\cdot\mathbf{z}_B$ will go through the sine activation function to obtain the final explicit formula-based common features, i.e., 
\begin{equation}
\label{manual_feature}
\begin{aligned}
\Upsilon_A=\sin\left(\bm{\varrho}^T\cdot\mathbf{z}_A\right),\\
\Upsilon_B=\sin\left(\bm{\varrho}^T\cdot\mathbf{z}_B\right).
\end{aligned}
\end{equation}
Given $\Upsilon_A\approx f_A\approx f_B\approx\Upsilon_B$, the secret keys leveraging these explicit-form features can be generated by adopting the standard key quantization \cite{6171198,mathur2008radio}, reconciliation \cite{10.1007/3-540-48285-7_35}, and privacy amplification \cite{impagliazzo1989pseudo}. One realization of the linear coefficient vector $\bm{\varrho}$ is provided in Appendix \ref{appendix_coe}, which will be used as the explicit-formula-based feature generator in the simulation part.}

\section{NN-based MITM-RIS Eve: the Worst-Case to Test}\label{MITM_RIS_Eve_NN}
In this section, we describe a strong NN-based MITM-RIS Eve, which will be used in the simulation section to test our designed PL-SKGs. We consider the worst scenario where this MITM-RIS Eve has the full knowledge of the trained common feature generator NNs, or the explicit forms in (\ref{manual_feature}). Here, even with the full knowledge of Alice's and Bob's feature generators, Eve cannot directly reconstruct their generated common features, as Eve does not know the exact send and receive signals of Alice and Bob. Yet, via the MITM-RIS, Eve can receive the signals from Alice and Bob, i.e., $\mathbf{y}_R^{(A)}$ and $\mathbf{y}_R^{(B)}$, which thereby poses the threat of recovering the legitimate common features via a well-trained NN.   

It is worth noting that this NN-based MITM-RIS Eve is conceptually different from the adversarial Mallory in our adversarial learning-based framework, as the latter is deliberately employed by legitimate Alice and Bob to ensure their features cannot be estimated by potential real Eves. 

\subsection{Structure of MITM-RIS Eve NN}
The aim of MITM-RIS Eve NN is to reconstruct the legitimate common features, via her received signals from Alice and Bob, i.e., $\mathbf{y}_R^{(A)}$ and $\mathbf{y}_R^{(B)}$, and the configurable phases $\mathbf{w}$. By denoting the parameters of MITM-RIS Eve NN as $\bm{\theta}_E$, its output is expressed as:
\begin{equation}
\label{eq22}
f_E=\Psi_{\bm{\theta}_E}\Big(\Re[\mathbf{y}_{R}^{(A)}], \Im[\mathbf{y}_{R}^{(A)}], \Re[\mathbf{y}_{R}^{(B)}],\Im[\mathbf{y}_{R}^{(B)}], \Re[\mathbf{w}], \Im[\mathbf{w}]\Big).
\end{equation}

The structure of MITM-RIS Eve NN consists of one input layer, 3 hidden layers, and an output layer. The input is RIS Eve's received signals from Alice and Bob, and the RIS phase vector. To further strengthen this Eve, the numbers of neurons of 3 hidden layers are assigned as 2048, 512, and 128, respectively, where the activation functions are ReLU. The output layer has 1 neuron that forms Eve's reconstructed features $f_E$.

\subsection{Training Dataset of MITM-RIS Eve}
To create the training dataset, an assumed Alice and Bob are simulated. The channel models between assumed Alice and Bob to MITM-RIS Eve and the combined RIS channels are modelled via (\ref{eq02}) and (\ref{eq1}). To distinguish with real Alice and Bob, these models are denoted as $\underline{\mathbf{g}}_{AR}$, $\underline{\mathbf{g}}_{BR}$, $\underline{g}_{AB}$, and $\underline{g}_{BA}$. Then, Alice's and Bob's send and receive signals are simulated as $\underline{x}_A,\underline{x}_B\in\{\sqrt{P_t}\exp(j\iota)|\iota\in[0,2\pi]\}$, and $\underline{y}_A,\underline{y}_B$ by taking the assumed channels into (\ref{eq3}). The received signals of MITM-RIS Eve are simulated as $\underline{\mathbf{y}}_R^{(A)}$ and $\underline{\mathbf{y}}_R^{(B)}$ by taking $\underline{\mathbf{g}}_{AR}$, $\underline{\mathbf{g}}_{BR}$ and $\underline{x}_A,\underline{x}_B$ into (\ref{eve_receive}). 

With the full knowledge of Alice's and Bob's common feature generators (trained NNs or the explicit forms), MITM-RIS Eve is able to generate common features of assumed Alice and Bob, denoted as $\underline{f}_A$ and $\underline{f}_B$, given the simulated send and receive signals of Alice and Bob, i.e., $\underline{x}_A,\underline{y}_A$ and $\underline{x}_B,\underline{y}_B$. 

As such, one training data can be expressed as $(\underline{\mathbf{y}}_R^{(A)},\underline{\mathbf{y}}_R^{(B)},\mathbf{w})$ with the label $(\underline{f}_A,\underline{f}_B)$. The total dataset for training the MITM-RIS Eve NN is the samples of $N=10^7$ pairs of data and labels.

\subsection{Supervised Training of MITM-RIS Eve NN}
To train this Eve NN to generate the highly correlated features with Alice and Bob, the loss function is designed by maximizing the absolute correlations between MITM-RIS Eve's output and simulated legitimate features, i.e., 
\begin{equation}
\label{eq24}
\text{Loss}_E(\bm{\theta}_E) = -\left|\rho(\underline{f}_E,\underline{f}_A)\right| - \left|\rho(\underline{f}_E,\underline{f}_B)\right|,
\end{equation}
where $\underline{f}_E$ is computed by taking the training data of $(\underline{\mathbf{y}}_R^{(A)},\underline{\mathbf{y}}_R^{(B)},\mathbf{w})$ into the NN model in (\ref{eq22}). 

To derive the trained parameter $\bm{\theta}_E$, Adam optimizer is adopted to minimize the loss function in (\ref{eq24}), with a learning rate of $10^{-5}$. We will show in the simulation section that even the MITM-RIS Eve, with its exact knowledge of the legitimate feature generators, it is still impossible to reconstruct the legitimate common features for secret key generation.

\section{Simulation Results}\label{sim}
In this section, we evaluate our trained and designed legitimate feature generators, to confront the trained MITM-RIS Eve NN. 

\subsection{Simulation Setup}\label{sim_set}
The model configuration is provided below. In a 3D space, the MITM-RIS Eve is located at $(0,0,0)$ with unit $m$, and the positions of Alice and Bob are within the $25$m from RIS. The direct channels from Alice and Bob to MITM-RIS Eve are modelled in (\ref{eq02}) according to \cite{9300189}, where a square structure of RIS is considered with $M=M_x\times M_y=40\times40=1600$ elements, and the amplifying power of the MITM-RIS Eve is $A_m=40$dB \cite{10100938}. The referenced path loss is set as $C_0=-30$dB at the reference distance (i.e., $1$m). The path loss exponents are $\alpha=3$. The number of multi-paths is $L=10$ with random half-space elevation and azimuth angles independently and randomly distributed over $\mathcal{U}[-\pi/2,\pi/2]$. The power of the sent signals from Alice and Bob is set as $P_t=0.1$W. The receiving noise variance for Alice, Bob and MITM-RIS Eve ranges from $\sigma^2\in[-115,-90]$dBW. 

\begin{figure}[!t]
\centering
\includegraphics[width=3.4in]{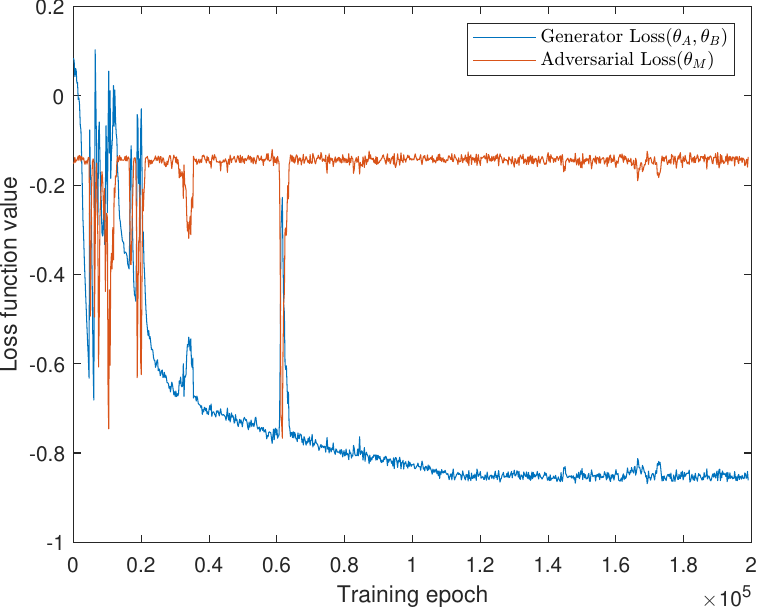}
\caption{Convergence of loss functions over training epochs for the adversarial learning-based legitimate common feature generator NNs and the adversarial NN.}
\label{figs1}
\end{figure}

\subsection{Performance of designed Adversarial Learning-based Common Feature Generators}
\begin{figure}[!t]
\centering
\includegraphics[width=3.4in]{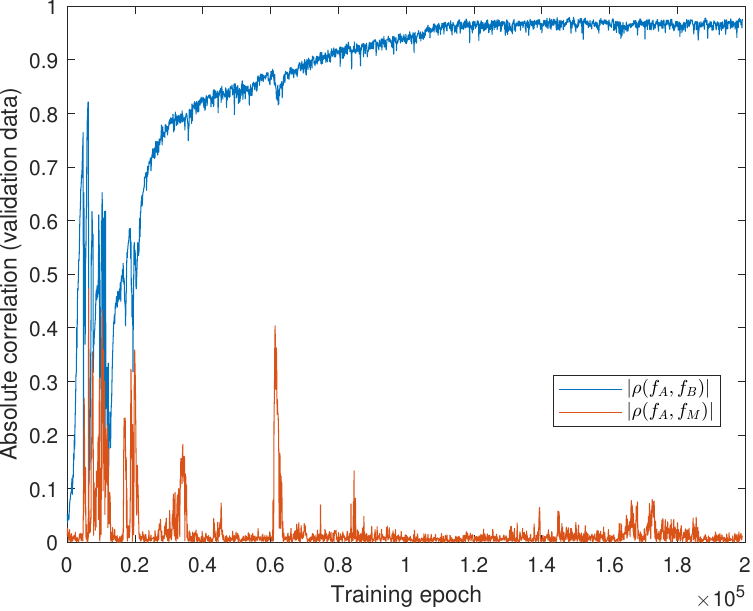}
\caption{Convergence (to $1$) of absolute correlation coefficients of common features generated by Alice's and Bob's feature generator NNs. }
\label{figs2}
\end{figure}

We first show the training and testing performance of our designed adversarial learning-based framework. It is shown in Fig.~\ref{figs1} that the loss function of legitimate feature generator NNs converges within $2\times10^5$ training epochs. According to (\ref{eq10}), this loss function contains to (i) maximize the absolute correlation coefficient between Alice's and Bob's features, and (ii) minimize the absolute correlation coefficient between Alice's (Bob's) and adversarial Mallory's features. These are shown to be achieved in Fig.~\ref{figs2}, where the absolute correlation coefficient between Alice's and Bob's features converges to $1$ and that between Alice's (Bob's) and adversarial Mallory's features converges to $0$. It is worth noting that during the training stage, some spikes appear in both Figs.~\ref{figs1} and \ref{figs2} (e.g., from $0$-$10^5$ epochs). This is due to the adversarial characteristics to jump out of the local optima area to better satisfy both legitimate and adversarial objectives in the generators' loss function. Such spikes then gradually disappear with the increase of epoch, which indicates the finding of the stable optimal area. As such, by combining the results from Figs. \ref{figs1} and \ref{figs2}, we demonstrate the successful training of the designed adversarial learning framework.

\begin{figure}[!t]
\centering
\includegraphics[width=3.4in]{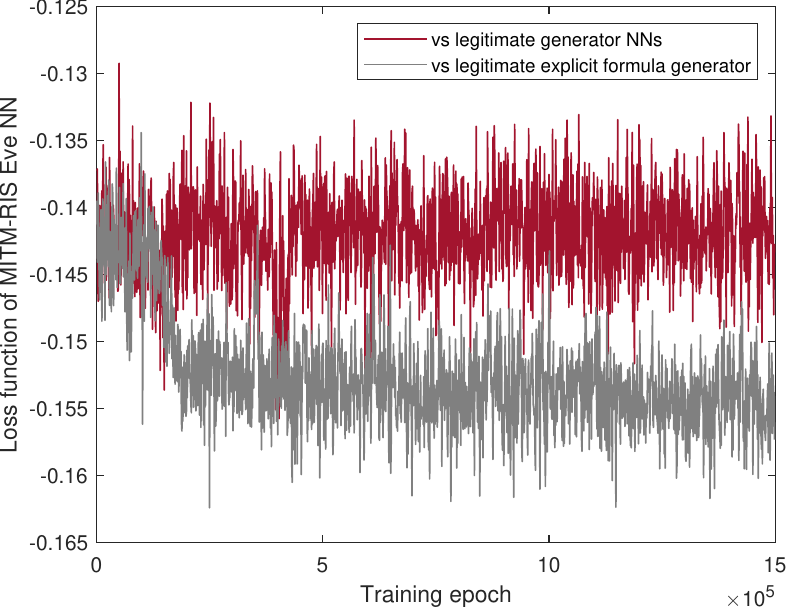}
\caption{Training performance of NN-based MITM RIS Eve: versus (v.s.) legitimate trained common feature generator NNs, and xAI induced explicit-form common feature generators.}
\label{figs3}
\end{figure}

We next present the results from training and validating the NN-based MITM-RIS Eve, described in Section \ref{MITM_RIS_Eve_NN}. Here, we evaluate the performance of the MITM-RIS Eve NN to learn the common features generated from both the trained legitimate generator NNs and the explicit formula-based generators in (\ref{mm1})-(\ref{manual_feature}). \textcolor{blue}{Fig.~\ref{figs3} demonstrates that the training of the MITM-RIS Eve NN fails to converge to its expected minimum value when attempting to learn either the legitimate common feature generator NN or the explicit formula-based common feature generators. According to the loss function in (\ref{eq24}), successful training of the MITM-RIS Eve NN would result in absolute correlation coefficients between Eve’s and Alice’s (or Bob’s) features approximating $1$, thereby rendering the loss function, which negatively sums these absolute correlation coefficients, approximating $-2$. However, even after $1.5\times10^{6}$ training epochs, Eve’s loss function remains between $-0.1$ and $-0.2$. This indicates that the NN-based MITM-RIS Eve is unable to effectively learn the common feature spaces from either the trained legitimate generator NNs or the explicit formula-based generators.}  

\begin{figure}[!t]
\centering
\includegraphics[width=3.4in]{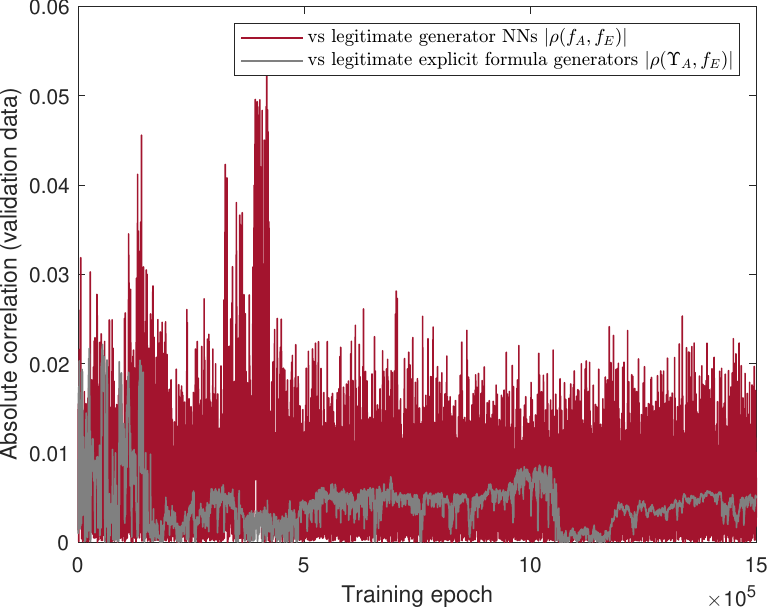}
\caption{The absolute correlation coefficients of features generated by the MITM-RIS Eve NN and Alice during Eve's training stage.}
\label{figs4}
\end{figure}

In Fig.~\ref{figs4}, we present the results for two absolute correlation coefficients between the MITM-RIS Eve's features and Alice's features, generated by the trained common feature generator NN and the explicit-formula-based generator, respectively: $|\rho(f_A,f_E)|$ and $|\rho(\Upsilon_A,f_E)|$. It is seen that even after $1.5\times10^{6}$ epochs, the NN-based MITM-RIS Eve cannot learn the legitimate features, as the absolute correlation coefficients do not exceed $0.1$. 

\begin{figure}[!t]
\centering
\includegraphics[width=3.4in]{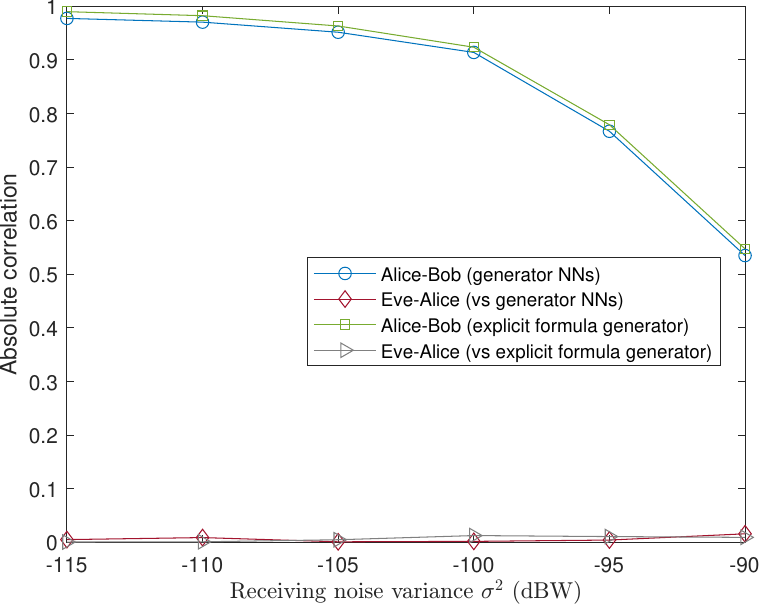}
\caption{The absolute correlation coefficients of features between Alice-Bob, and Alice-Eve, under different levels of receiving noise variance $\sigma^2$.}
\label{figs5}
\end{figure}

We next evaluate our proposed adversarial learning-based and explicit-formula-based common feature generators under different receiving noise variance $\sigma^2\in[-115, -90]$dBW. In Fig.~\ref{figs5}, we provide the absolute correlation coefficients of features between (i) Alice and Bob using feature generator NNs, (ii) Alice and Bob using the explicit formula-based feature generator, (iii) NN-based MITM-RIS Eve and Alice using feature generator NN, and (iv) NN-based MITM-RIS Eve and Alice using explicit formula-based feature generator. It is seen that correlation coefficients between Alice's and Bob's features are from $1$ to $0.5$ as the noise variance increases from $-115$dBW to $-90$dBW, as opposed to that the correlation coefficients between MITM-RIS Eve's and Alice's features remain below $0.02$ throughout. By combining the results from Figs. \ref{figs3}-\ref{figs5}, we demonstrate the effectiveness of the proposed adversarial learning-based approach, along with the xAI-induced explicit-formula-based common feature generator, in defending against the deep learning-based MITM-RIS Eve.

\begin{figure}[!t]
\centering
\includegraphics[width=3.4in]{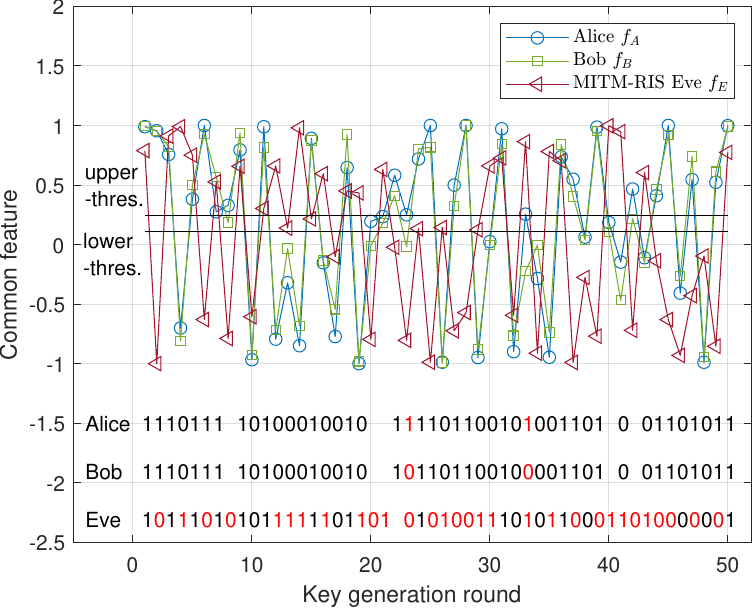}
\caption{Illustration of secret key quantization from common features by upper and lower thresholds.}
\label{figa1}
\end{figure}

\subsection{Performance of Secret Keys}
After the generation of the common features at Alice and Bob, they will individually quantize these into binary keys. In this part, we evaluate the secret keys quantized from the common features, by analyzing the key agreement rates of our proposed common feature generators, versus different levels of receiving noise variance $\sigma^2$. The secret key quantization technique used here is the standard two-threshold method, i.e., \cite{mathur2008radio}
\begin{equation}
\label{key_quant}
k_u=
\begin{cases}
1 & f_{u}>\gamma_{u}^{(+)}\\
0 & f_{u}<\gamma_{u}^{(-)}
\end{cases}
,~u\in\{A,B,E\}.
\end{equation}
where $k_A,k_B,k_E\in\{0,1\}$ are the quantized binary keys by Alice, Bob and MITM-RIS Eve using their features $f_u$, and $\gamma_{u}^{(\pm)}$ are the upper and lower quantization thresholds, assigned as $\gamma_{u}^{(\pm)}\triangleq\mathbb{E}\left(f_u\right)\pm\gamma\cdot \mathbb{D}\left(f_u\right)$ with $\gamma=0.1$. In each key generation round, the key will be quantized as $1$ if the current feature is larger than the predefined upper threshold, or as $0$ if it falls below the predefined lower threshold. If the feature lies within the region between the two thresholds, it will be discarded for key generation. We observe from Fig.~\ref{figa1} that leveraging the highly correlated common features of Alice and Bob provided by our proposed scheme and is non-learnable by MITM-RIS Eve, the quantized secret keys of Alice and Bob present a high agreement rate as opposed to the MITM-RIS Eve generated keys. 

\begin{figure}[!t]
\centering
\includegraphics[width=3.4in]{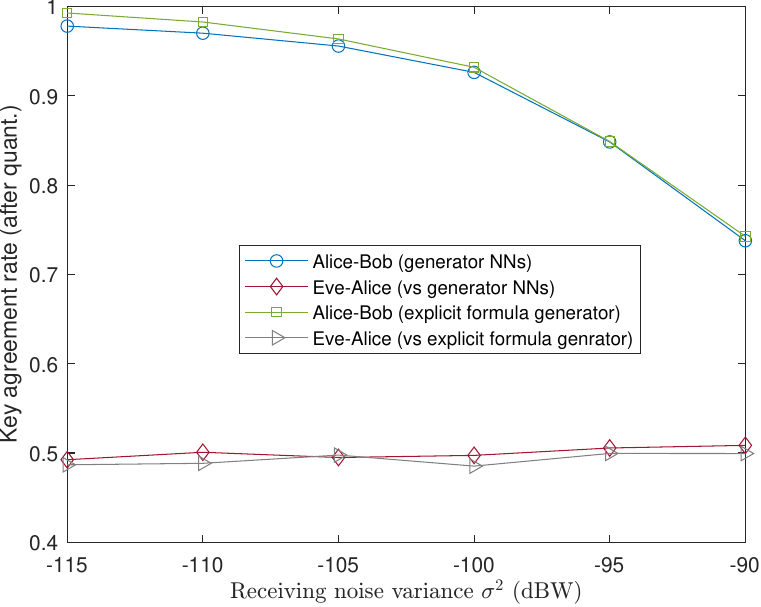}
\caption{Key agreement rate after quantization from Alice's and Bob's common features. }
\label{figs6}
\end{figure}

We then provide in Fig.~\ref{figs6} the secret key agreement rates of our proposed adversarial learning-based and its xAI-induced explicit formula-based common feature generators, under different levels of receiving noise variance, i.e., $\sigma^2\in[-115,-90]$dBW. It is seen that with the decrease of the receiving noise variance $\sigma^2$, the key agreement rates between Alice and Bob approach $1$, while such rates between the MITM-RIS Eve and the legitimate parties remain $0.5$ (the random guess probability), which corresponds to the correlation coefficient results in Fig.~\ref{figs5}. As such, combining the evaluations from Figs.~\ref{figs5}-\ref{figs6}, our proposed adversarial learning-based scheme and its xAI-induced explicit form-based common features, show promising capabilities in generating secret keys to countermeasure the MITM-RIS Eve threat.

\subsection{Performance Comparison with Existing Works}\label{compare_daae}
In this part, we will compare (from concepts to results) our proposed adversarial learning-based and xAI-induced common features, with (i) the other adversarial learning-based scheme, named DAAE \cite{zhou2023physical}, and (ii) the non-adversarial common feature NN \cite{he2022deep}.

Indeed, the idea of adversarial learning has also been adopted by the work in \cite{zhou2023physical}, where the DAAE scheme was proposed to deal with the traditional Eve that shares part of CSIs with legitimate users (e.g. when Eve is located within the half-wavelength of one legitimate user). Leveraging the CSI differences between legitimate parties and Eve, DAAE encodes Alice's and Bob's reciprocal CSIs into common features that cannot be decoded by a potential Eve. According to \cite{zhou2023physical}, the whole structure of DAAE requires (i) one legitimate encoder NN, denoted as $\text{En}_{\bm{\vartheta}}(\cdot)$, to convert the reciprocal CSIs of Alice and Bob to common features, (ii) one legitimate decoder NN, denoted as $\text{De}_{\bm{\vartheta}_{L}}(\cdot)$, to convert the common features to reciprocal CSIs, and (iii) an adversarial decoder NN, denoted as $\text{De}_{\bm{\vartheta}_{E}}(\cdot)$ that converts common features to Eve's estimated CSI. As such, the loss functions of the encoder (legitimate) and decoders (legitimate and adversarial) are \cite{zhou2023physical}:
\begin{equation}
\label{compa}
\begin{aligned}
\text{Loss}&(\bm{\vartheta},\bm{\vartheta}_{L})=\text{MSE}\left[\text{De}_{\bm{\vartheta}_{L}}\left(\text{En}_{\bm{\vartheta}}(\hat{h}_A)\right),\hat{h}_B\right]\\&-\text{MSE}\left[\text{De}_{\bm{\vartheta}_{E}}\left(\text{En}_{\bm{\vartheta}}(\hat{h}_A)\right),\hat{h}_E\right]\\
\text{Loss}&(\bm{\vartheta}_{E})=\text{MSE}\left[\text{De}_{\bm{\vartheta}_{E}}\left(\text{En}_{\bm{\vartheta}}(\hat{h}_A)\right),\hat{h}_E\right].
\end{aligned}
\end{equation}
After simultaneously minimizing the two loss functions, the trained encoder $\text{En}_{\bm{\vartheta}}(\cdot)$ will be used as the DAAE-based common feature generators for Alice and Bob.

It can be observed from (\ref{compa}) that DAAE cannot be implemented for two-way signal-based legitimate common feature generation. This is because the loss functions in (\ref{compa}) require the existence of an inverse mapping from the common features to the input of the encoder. Without this, the decoder NN cannot be trained successfully. However, this condition becomes difficult to satisfy when using two-way signals for feature generation, as inversely deriving both $(x_A,y_A)$ and $(x_B,y_B)$ from $f_A\approx f_B$ with independent $x_A$ and $x_B$ poses a significant challenge. In contrast, our proposed adversarial learning-based schemes do not require the inverse mapping from the common feature space to the two-way signal space. This therefore paves the way to finding the common feature space from two-way signals to defend the MITM-RIS Eve.

\begin{figure}[!t]
\centering
\includegraphics[width=3.4in]{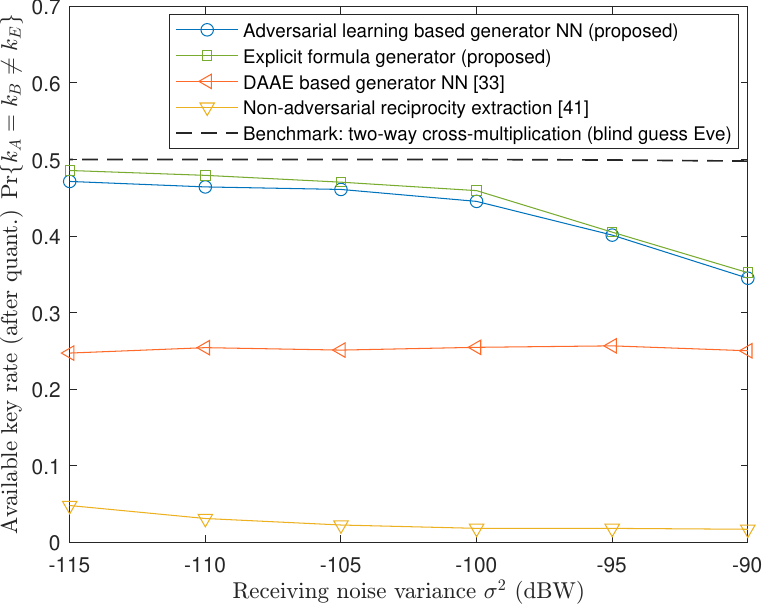}
\caption{Performance comparison of generated secret keys among our proposed adversarial learning-based and the xAI deduced explicit-formula-based common feature generators, the non-adversarial reciprocity learning NN \cite{he2022deep}, and the existing DAAE-based common feature generator \cite{zhou2023physical}.}
\label{fig_add}
\end{figure}

{\color{blue}
We compare the secret key performances among the proposed adversarial learning and its xAI-induced explicit-formula-based common feature generators, the DAAE-based common feature generators \cite{zhou2023physical}, and the non-adversarial reciprocity extraction NN \cite{he2022deep}. In Fig. \ref{fig_add}, the available key rate (y-axis) is adopted as the metric to assess the ratio of matched keys (after the quantization process) between Alice and Bob that cannot be recovered by Eve, i.e., $\text{Pr}\{k_A=k_B\neq k_E\}$ \cite{10100938,9954393}. We first plot the benchmark available key rate through the two-way cross-multiplication-based common feature in (\ref{cross_multi}), in the face of a random blind guess Eve. It is seen from Fig. \ref{fig_add} that $\text{Pr}\{k_A=k_B\neq k_E\}=0.5$, as the key agreement rate between Alice and Bob approach $\text{Pr}\{k_A=k_B\}\approx1$, and the blind guess Eve can present a probability of $\text{Pr}\{k_A=k_E\}\approx0.5$ to guess every single binary key. 

We then compare available secret key rates of the proposed adversarial learning-based schemes with non-adversarial schemes. Here, we consider the non-adversarial reciprocity extraction NN from \cite{he2022deep}, as it is compatible with the two-way signal input. It is seen from Fig. \ref{fig_add} that the available key rate of the non-adversarial scheme is less than $0.1$. This is because even if a high key agreement rate between Alice and Bob can be achieved, i.e., $\text{Pr}\{k_A=k_B\}\approx1$, the MITM-RIS Eve can learn the legitimate Alice's and Bob's common features, given the lack of adversarial part considered in the legitimate training process, thereby making $\text{Pr}\{k_A=k_E\}\approx1$ and $\text{Pr}\{k_A=k_B\neq k_E\}<0.1$. 

Next, it is observed that the proposed adversarial learning-based, and its explicit-formula-based common feature generators, can achieve higher available key rates, as opposed to the existing DAAE. As explained in the above analysis under (\ref{compa}), this is due to the lack of inverse mapping from its encoder-generated feature to the input two-way signals (i.e., from $f_A\approx f_B$ back to $(x_A,y_A)$ and $(x_B,y_B)$ with independent sending $x_A$ and $x_B$ of Alice and Bob). This further causes the unsuccessful learning of common features, which leads to $\text{Pr}\{k_A=k_B\}\approx0.5$ and $\text{Pr}\{k_A=k_E\}\approx0.5$, resulting in $\text{Pr}\{k_A=k_B\neq k_E\}\approx0.25$.  
By contrast, our proposed adversarial learning-based schemes do not require the inverse mapping from the common feature space to the two-way signal space. Leveraging the theoretical mutual information gap deduced in (\ref{theo1}), the proposed schemes are capable of exploiting the explainable polynomial and periodic feature bettering to identify the common feature space that cannot be learned by MITM-RIS Eve.}

\section{Conclusion}
This paper proposes an adversarial learning framework to address MITM-RIS Eve security challenges. The proposed approach leverages adversarial learning to generate common features through legitimate two-way signals that embrace reciprocal CSIs. By designing the adversarial loss function to embrace both the correlations between Alice's and Bob's generator outputs, as well as correlations between Alice's outputs and the adversarial Mallory NN, Alice and Bob are able to learn the common feature space that cannot be learned by MITM-RIS Eve. Furthermore, the use of the Meijer G function-based symbolic representation facilitates the design of explicit formula-based legitimate common feature generators. Simulation results demonstrate the efficiency of the proposed adversarial learning-based and symbolic-based PL-SKGs, showcasing high key agreement rates (above $0.9$ with receiving SNR>$15$dB) between legitimate users. Notably, these solutions demonstrate resilience against the MITM-RIS Eve, \textcolor{blue}{even when it possesses the full knowledge of the legitimate feature generators} (NNs or the explicit formulas). This promising outcome signifies a significant step toward establishing secure wireless communications in the context of untrusted or malicious reflective devices, laying the groundwork for future 6G networks.

\appendices

{\color{blue}
\section{Deduction of Positive SKR in (\ref{theo1})}\label{appendix1}
To simplify the notation, we denote the reciprocal channel between Alice and Bob as $g=g_{AB}=g_{BA}$, which follows the complex Gaussian distribution with variance $\sigma_g^2$ \cite{10100938}. Then, we show the proof of the first part of (\ref{theo1}), i.e., 
\begin{equation}
\label{s1}
I\left(
x_A,y_A;
x_B,y_B
\right)-I\left(
x_A,y_A;
\mathbf{y}_R^{(A)},\mathbf{y}_R^{(B)},\mathbf{w}
\right)>0,
\end{equation}
which is equivalent to the second part by replacing $x_A,y_A$ by $x_B,y_B$. 

The first mutual information term of (\ref{s1}) can be expressed in terms of the entropy $h(x)=-\int p(x)\log_2p(x)dx$, i.e.,  
\begin{equation}
\label{s2}
\begin{aligned}
&I\left(
x_A,y_A;
x_B,y_B
\right)
\\
=&
h\left(
x_A,y_A\right)
+
h\left(
x_B,y_B\right)
-
h\left(
x_A,y_A,x_B,y_B\right)\\
=&h(x_A)\!\!+\!\!h(y_A)\!\!+\!\!h(x_B)\!\!+\!\!h(y_B)\!\!-\!\!h(y_A,y_B|x_A,x_B)\!\!-\!\!h(x_A,x_B)\\
=&h(y_A)+h(y_B)-h(y_A,y_B|x_A,x_B),
\end{aligned}
\end{equation}
given the independent pairs of ($x_A$, $y_A=g\cdot x_B+n_A$), ($x_B$, $y_B=g\cdot x_A+n_B$), and $(x_A,x_B)$. Similarly, the second mutual information term of (\ref{s1}) is further expressed as:
\begin{equation}
\label{s3}
\begin{aligned}
&I\left(
x_A,y_A;
\mathbf{y}_R^{(A)},\mathbf{y}_R^{(B)},\mathbf{w}\right)\\
=&h\left(
x_A,y_A\right)+
h\left(
\mathbf{y}_R^{(A)},\mathbf{y}_R^{(B)},\mathbf{w}
\right)-
h\left(
x_A,y_A,\mathbf{y}_R^{(A)},\mathbf{y}_R^{(B)},\mathbf{w}\right)\\
=&h(y_A)+h\left(\mathbf{y}_R^{(A)}\right)-h\left(y_A,\mathbf{y}_R^{(A)}|x_A,\mathbf{y}_R^{(B)},\mathbf{w}\right).
\end{aligned}
\end{equation}
As such, the mutual information difference in (\ref{s1}) is:
\begin{equation}
\label{s4}
\begin{aligned}
&I\left(
x_A,y_A;
x_B,y_B
\right)-I\left(
x_A,y_A;
\mathbf{y}_R^{(A)},\mathbf{y}_R^{(B)},\mathbf{w}
\right)\\
=&h(y_B)\!\!-\!\!h\!\left(\!\mathbf{y}_R^{(A)}\!\right)\!\!+\!\!h\!\left(\!y_A,\mathbf{y}_R^{(A)}|x_A,\mathbf{y}_R^{(B)}\!,\!\mathbf{w}\!\right)\!\!-\!\!h(y_A,y_B|x_A,x_B),
\end{aligned}
\end{equation}

The first term of the right-hand side of (\ref{s4}) is computed as:
\begin{equation}
\label{s5}
h(y_B)=h(gx_A+n_B)=0.5\log_2\left[2\pi e\left(P_t\sigma_g^2+\sigma^2\right)\right],
\end{equation}
given $\mathbb{D}(gx_A+n_B)=|x_A|^2\sigma_g^2+\sigma^2=P_t\sigma_g^2+\sigma^2$. 

The second term of the right-hand side of (\ref{s4}) is computed as:
\begin{equation}
\label{s6}
\begin{aligned}
&h\left(\mathbf{y}_R^{(A)}\right)=h\left(\mathbf{g}_{AR}x_A+\mathbf{n}_R^{(A)}\right)\\
=&0.5\log_2\left[(2\pi e)^M\cdot\text{det}\left(P_t\Sigma_{AR}+\sigma^2\mathbf{I}_M\right)\right],
\end{aligned}
\end{equation}
given $\mathbb{D}(\mathbf{g}_{AR}x_A+\mathbf{n}_R^{(A)})=P_t\bm{\Sigma}_{AR}+\sigma^2\mathbf{I}_M$, with $\bm{\Sigma}_{AR}$ the covariance matrix of complex Gaussian channel $\mathbf{g}_{AR}$. 

To compute the third term of the right-hand side of (\ref{s4}), i.e., $h(y_A,\mathbf{y}_R^{(A)}|x_A,\mathbf{y}_R^{(B)},\mathbf{w})$, we compute the covariance matrix of the conditional complex Gaussian variables $y_A,\mathbf{y}_R^{(A)}|x_A,\mathbf{y}_R^{(B)},\mathbf{w}$. By rewriting $y_A=\sum_m w_m g_{AR,m}(g_{BR,m}x_B)+n_A$, we obtain the covariance matrix of $y_A,\mathbf{y}_R^{(A)}$, conditioned on $x_A,\mathbf{y}_R^{(B)},\mathbf{w}$, i.e., 
\begin{equation}
\label{s7}
\begin{aligned}
&\text{Cov}\left(y_A,\mathbf{y}_R^{(A)}|x_A,\mathbf{y}_R^{(B)},\mathbf{w}\right)\\
=&
\begin{bmatrix}
\bm{\zeta}^H\bm{\Sigma}_{AR}\bm{\zeta}+\sigma^2 & x_A^*\bm{\zeta}^T\bm{\Sigma}_{AR}^H\\
x_A\bm{\Sigma}_{AR}\bm{\zeta} & P_t\bm{\Sigma}_{AR}+\sigma^2\mathbf{I}_M,
\end{bmatrix}
\end{aligned}
\end{equation}
where $\bm{\zeta}=\mathbf{y}_R^{(B)}\circ\mathbf{w}=x_A\mathbf{g}_{AR}\circ\mathbf{w}+\mathbf{n}_R^{(A)}\circ\mathbf{w}\sim\mathcal{CN}(0,\sigma_\zeta^2\mathbf{I}_M)$, with 
\begin{equation}
\label{s_7_5}
\sigma_\zeta^2=P_tA_EC_0d_{AR}^{-\alpha_L}+A_E\sigma^2,
\end{equation}
given the fact that $\mathbb{E}(g_{AR,m}w_mg_{AR,m'}^*w_m^*)=0$ when $m\neq m'$, and  $\mathbb{E}(g_{AR,m}w_mg_{AR,m'}^*w_m^*)=A_EC_0d_{AR}^{-\alpha_L}$ \cite{10100938} when $m=m'$. 
Then, the entropy of $y_A,\mathbf{y}_R^{(A)}|x_A,\mathbf{y}_R^{(B)},\mathbf{w}$ is deduced as:
\begin{equation}
\label{s8}
\begin{aligned}
&h\left(y_A,\mathbf{y}_R^{(A)}|x_A,\mathbf{y}_R^{(B)},\mathbf{w}\right)\\
=&0.5\mathbb{E}_{\bm{\zeta}}\log_2\left[(2\pi e)^{(M+1)}\text{det}\left(\text{Cov}\left(y_A,\mathbf{y}_R^{(A)}|x_A,\mathbf{y}_R^{(B)},\mathbf{w}\right)\!\right)\!\right].
\end{aligned}
\end{equation}
The determinant in (\ref{s8}) can be further simplified as: 
\begin{equation}
\label{s9}
\begin{aligned}
&\text{det}\left(\text{Cov}\left(y_A,\mathbf{y}_R^{(A)}|x_A,\mathbf{y}_R^{(B)},\mathbf{w}\right)\right)\\
=&\left[\bm{\zeta}^H\bm{\Sigma}_{AR}\bm{\zeta}\!+\!\sigma^2\!-\!P_t\bm{\zeta}^H\bm{\Sigma}_{AR}\!\left(P_t\bm{\Sigma}_{AR}+\sigma^2\mathbf{I}_M\right)^{-1}\!\!\bm{\Sigma}_{AR}\bm{\zeta}\right]\\
&\cdot\text{det}\left(P_t\bm{\Sigma}_{AR}+\sigma^2\mathbf{I}_M\right).
\end{aligned}
\end{equation}
We then simplify the first term of the right-hand side of (\ref{s9}), which is:
\begin{equation}
\label{s10}
\begin{aligned}
&\bm{\zeta}^H\bm{\Sigma}_{AR}\bm{\zeta}\!+\!\sigma^2\!-\!P_t\bm{\zeta}^H\bm{\Sigma}_{AR}\!\left(P_t\bm{\Sigma}_{AR}+\sigma^2\mathbf{I}_M\right)^{-1}\!\!\bm{\Sigma}_{AR}\bm{\zeta}\\
\overset{(a)}{=}&\bm{\zeta}^H\mathbf{U}\bm{\Lambda}\mathbf{U}^H\bm{\zeta}\!+\!\sigma^2\!-\!P_t\bm{\zeta}^H\mathbf{U}\bm{\Lambda}\!\left(P_t\bm{\Lambda}+\sigma^2\mathbf{I}_M\right)^{-1}\!\!\bm{\Lambda}\mathbf{U}^H\bm{\zeta}\\
\overset{(b)}{=}&\bm{\zeta}^H\mathbf{U}\text{diag}\left(\left[\frac{\lambda_1\sigma^2}{P_t\lambda_1+\sigma^2},\cdots,\frac{\lambda_M\sigma^2}{P_t\lambda_M+\sigma^2}\right]\right)\mathbf{U}^H\bm{\zeta}\!+\!\sigma^2\\
\overset{(c)}{=}&\sigma_\zeta^2\cdot\text{tr}\!\!\left\{\!\mathbf{U}\text{diag}\left(\left[\frac{\lambda_1\sigma^2}{P_t\lambda_1+\sigma^2},\cdots,\frac{\lambda_M\sigma^2}{P_t\lambda_M+\sigma^2}\right]\right)\mathbf{U}^H\!\right\}\!\!+\!\!\sigma^2\\
\overset{(d)}{=}&\sigma_\zeta^2\sum_{m=1}^M\frac{\lambda_m\sigma^2}{P_t\lambda_m+\sigma^2}+\sigma^2
\end{aligned}
\end{equation}
In (\ref{s10}), $(a)$ is by considering the eigen-decompose of the positive definite covariance matrix $\bm{\Sigma}_{AR}=\mathbf{U}\bm{\Lambda}\mathbf{U}^H$ with $\bm{\Lambda}=\text{diag}([\lambda_1,\cdots,\lambda_M])$, and $(P_t\bm{\Sigma}_{AR}+\sigma^2\mathbf{I}_M)^{-1}=(P_t\mathbf{U}\bm{\Lambda}\mathbf{U}^H+\sigma^2\mathbf{U}\mathbf{U}^H)^{-1}=\mathbf{U}(P_t\bm{\Lambda}+\sigma^2\mathbf{I}_M)^{-1}\mathbf{U}^H$. $(b)$ is to process the inverse of diagonal matrix. $(c)$ is due to that the element of $\bm{\zeta}$ independent and identically distributed, and that the number of element (i.e., the number of RIS elements $M$) is large (e.g., $M=1600$ in our work). $(d)$ is the trace computation. As such, by taking (\ref{s10}) back to (\ref{s9}) and (\ref{s8}), we express the third term of the right-hand side of (\ref{s4}), i.e., $h(y_A,\mathbf{y}_R^{(A)}|x_A,\mathbf{y}_R^{(B)},\mathbf{w})$, as:
\begin{equation}
\label{s11}
\begin{aligned}
&h\left(y_A,\mathbf{y}_R^{(A)}|x_A,\mathbf{y}_R^{(B)},\mathbf{w}\right)\\
=&0.5(M+1)\log_2(2\pi e)+0.5\log_2\text{det}\left(P_t\bm{\Sigma}_{AR}+\sigma^2\mathbf{I}_M\!\right)\\
&+0.5\log_2\left(\sigma_\zeta^2\sum_{m=1}^M\frac{\lambda_m\sigma^2}{P_t\lambda_m+\sigma^2}+\sigma^2\right)
\end{aligned}
\end{equation}

The last term of the right-hand side of (\ref{s4}), i.e., $h(y_A,y_B|x_A,x_B)$ can be computed by the covariance matrix of its conditional Gaussian variable $y_A,y_B|x_A,x_B$, which is:
\begin{equation}
\label{s12}
\text{Cov}\left(y_A,y_B|x_A,x_B\right)=
\begin{bmatrix}
P_t\sigma_g^2+\sigma^2 & x_Ax_B^*\sigma_g^2\\
x_A^*x_B\sigma_g^2 & P_t\sigma_g^2+\sigma^2
\end{bmatrix}.
\end{equation}
Leveraging (\ref{s12}), $h(y_A,y_B|x_A,x_B)$ is expressed as:
\begin{equation}
\label{s13}
\begin{aligned}
&h(y_A,y_B|x_A,x_B)\\
=&0.5\mathbb{E}_{x_A,x_B}\log_2\left[(2\pi e)^2\text{det}\left(\text{Cov}\left(y_A,y_B|x_A,x_B\right)\right)\right]\\
=&0.5\log_2\left[(2\pi e)^2+\sigma^2\left(2P_t\sigma_g^2+\sigma^2\right)\right].
\end{aligned}
\end{equation}

By taking (\ref{s5}), (\ref{s6}), (\ref{s11}) and (\ref{s13}) back into (\ref{s4}), the mutual information difference is further expressed as:
\begin{equation}
\label{s14}
\begin{aligned}
&I\left(
x_A,y_A;
x_B,y_B
\right)-I\left(
x_A,y_A;
\mathbf{y}_R^{(A)},\mathbf{y}_R^{(B)},\mathbf{w}
\right)\\
=&0.5\log_2(P_t\sigma_g^2+\sigma^2)+0.5\log_2\left(\sigma_\zeta^2\sum_{m=1}^M\frac{\lambda_m}{P_t\lambda_m+\sigma^2}+1\right)\\
&-0.5\log_2(2P_t\sigma_g^2+\sigma^2).
\end{aligned}
\end{equation}
We then analyze the second term of the right-hand side of (\ref{s14}), i.e., 
\begin{equation}
\label{s15}
\begin{aligned}
&\sigma_\zeta^2\sum_{m=1}^M\frac{\lambda_m}{P_t\lambda_m+\sigma^2}\overset{(\ref{s_7_5})}{>}A_EC_0d_{AR}^{-\alpha}\cdot\sum_{m=1}^M\frac{P_t\lambda_m}{P_t\lambda_m+\sigma^2}\\
=&A_EC_0d_{AR}^{-\alpha}M\cdot\overline{\eta}>A_EC_0d_{AR}^{-\alpha}M\cdot0.5,
\end{aligned}
\end{equation}
where $\overline{\eta}\rightarrow1$ when $P_t\lambda_m\gg\sigma^2$, and we loose the equality by $\overline\eta=0.5$ given $\sigma_2<-90$dBW and $P_t=-10$dBW. 

Next, according to the MITM-RIS Eve design shown in (6), (8) and Fig. 2 of \cite{10100938}, a MITM-RIS Eve works for any Alice and Bob within a distance, i.e., $d_{AR},d_{BR}\leq d$ should ensure $A_EMC_0d^{-2\alpha}>100d_{AB}^{-\alpha}$, in order to override the direct Alice-Bob channel with the deceptive Alice-RIS-Bob channel. This suggests $A_EMC_0d_{AR}^{-\alpha}>A_EMC_0d^{-\alpha}>100(d/d_{AB})^\alpha>100[d/(2d)]^\alpha$, which, taking into (\ref{s15}), provides:
\begin{equation}
\label{s16}
\sigma_\zeta^2\sum_{m=1}^M\frac{\lambda_m}{P_t\lambda_m+\sigma^2}>A_EC_0d_{AR}^{-\alpha}M\cdot0.5>\frac{0.5\cdot 100}{2^\alpha}>1,
\end{equation}
for any path-loss exponential factor $\alpha\in[2,4]$. As such, by taking (\ref{s16}) back into (\ref{s14}), we prove the positive mutual information gap (i.e., positive SKR) in (\ref{theo1}). 
}

{\color{blue}
\section{Trade-off Coefficient in Adversarial Loss Function Designs}\label{trade_off}
The trade-off coefficient $\lambda=0.8$ in the loss function (\ref{eq10}) is an empirical setting. We also plot the training performance with other $\lambda=0.2,0.4,0.6,1$. 

\begin{figure*}[!t]
\centering
\includegraphics[width=7in]{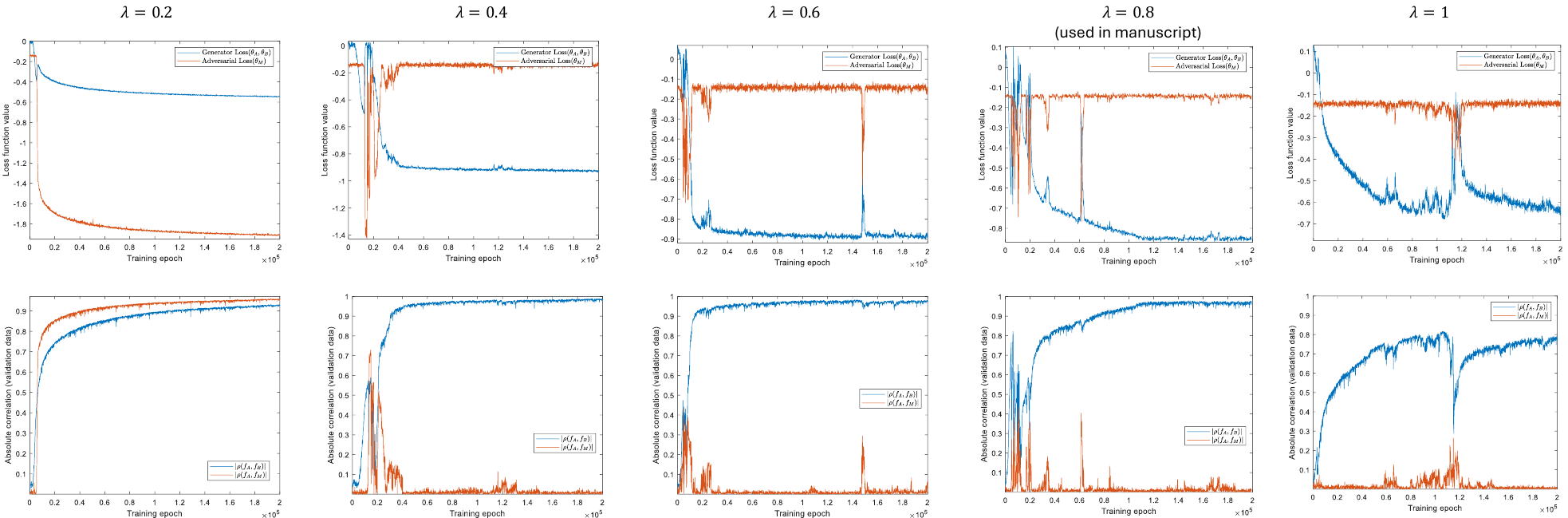}
\caption{\small{Training performance comparison with different trade-off adversarial parameters, $\lambda=0.2,0.4,0.6,0.8,1$.}}
\label{fig_trade_off}
\end{figure*}

Fig. \ref{fig_trade_off} shows the comparison of training performance with different trade-off parameters $\lambda$ set to operate the adversarial learning. It is shown that a small trade-off parameter $\lambda=0.2$ is not sufficient to take into account the adversarial MITM attack-related aspect, given its inability to find common features that the adversarial Mallory cannot learn (the correlation between legitimate feature and Mallory's feature approach $0.9$). 

Then, we find that with a large trade-off parameter like $\lambda=1$, it is true that Alice and Bob can learn some common feature space that cannot be learned from adversarial Mallory. However, the large portion of adversarial aspects in Alice's and Bob's loss functions affects their training performance, resulting in their learned common features being less correlated (the correlation coefficient is less than $0.9$). 

Next, for the trade-off parameters $\lambda=0.4,0.6,0.8$, the training performances are similar, where Alice and Bob can learn common feature spaces that cannot be learned from the adversarial part. In the manuscript, we use the result with the trade-off parameter as $\lambda=0.8$. }

\begin{figure}[!t]
\centering
\includegraphics[width=3.4in]{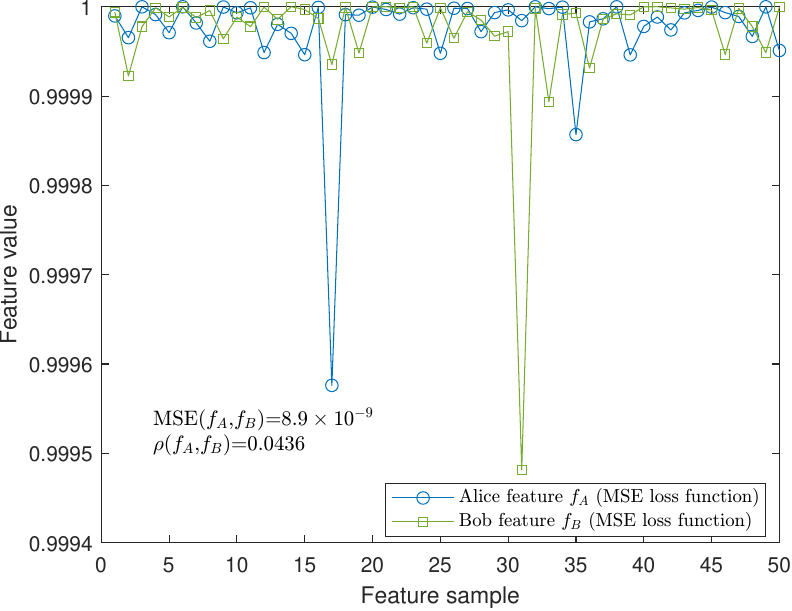}
\caption{Trained Alice's and Bob's common features via MSE-based loss function.}
\label{figsadd1}
\end{figure}

{\color{blue}
\section{Feature Learned by MSE Loss Function}\label{appendixC}
We provide in Fig. \ref{figsadd1} the trained Alice and Bob common features, by replacing the loss function in (\ref{eq10}) with the MSE-based loss function, i.e., 
\begin{equation}
\label{mse}
\begin{aligned}
&\text{Loss}(\bm{\theta}_A,\bm{\theta}_B)\\
=&\text{MSE}(f_A, f_B)-\lambda \left[\text{MSE}(f_A,f_M)+\text{MSE}(f_B,f_M)\right].
\end{aligned}
\end{equation}
It is shown in Fig. \ref{figsadd1} that indeed the MSE of features generated from Alice and Bob is small, which is $\text{MSE}(f_A,f_B)=8.9\times10^{-9}$. However, the correlation coefficient between Alice's and Bob's features is as low as $\rho(f_A,f_B)=0.0436$, which, as shown in Fig. \ref{figsadd1}, is not sufficient to generate secret keys. 
}

{\color{blue}
\section{Realization of Explicit-Formed Common Feature Generator}\label{appendix_coe}
Given the simulation parameters assigned in Section \ref{sim_set}, we provide one realization of the polynomial coefficient vector $\bm{\varrho}$ to construct the explicit-formula-based common feature generator in (\ref{mm1})-(\ref{manual_feature}), i.e.,  
\begin{equation}
\begin{aligned}
\bm{\varrho}=[&0,0.0330,-0.0104,0,0,0,0,-0.0207,-0.0531,0,\\
&-13.0482,0,-19.6748,0,0.0457,-0.0117,-0.0302,\\
&0.0430,-0.1009,-0.0136,0.1295,0,-0.3157,-0.0161,\\
&-0.3361,0.0172,0.0366,0,-19.7018,0,13.1063,0,\\
&-0.0264,-0.0119,0.0605,0,0,0,-0.3254,0.0261,\\
&-0.9464,-0.0246,0.3199,0,0,0,0.0301,0,0,0,\\
&0.0106,0.0128,-0.0469,0.0153,-0.1054,-0.0153,\\
&-0.3372,0,0.3190,0.0205,0.1345,0,0.0309,0,0,\\
&0,0.0102,0,-0.0250,0,0.0298,0,0,0,0,-0.0143,\\
&0,0.0167,0,0,-0.0250]. 
\end{aligned}
\end{equation}
This explicit-formula-based common feature generator for Alice and Bob is assessed in Simulations. 
}

\textcolor{blue}{It is also noteworthy that multiple formulas exist for common feature generators, each derived from different training outcomes of the common feature generator NNs. Each formula represents a unique learned common feature space that remains inaccessible to MITM-RIS.}

\bibliographystyle{IEEEtran}
\bibliography{myref.bib}

\end{document}